\documentclass[fleqn,usenatbib]{mnras}

\usepackage{newtxtext,newtxmath}

\usepackage[T1]{fontenc}
\usepackage[dvipsnames]{xcolor}
\usepackage{makecell}
\usepackage{longtable}

\DeclareRobustCommand{\VAN}[3]{#2}
\let\VANthebibliography\thebibliography
\def\thebibliography{\DeclareRobustCommand{\VAN}[3]{##3}\VANthebibliography}

\usepackage{graphicx}	
\usepackage{amsmath}	
\usepackage{multicol}
\usepackage{newtxtext,newtxmath}

\usepackage{caption,subcaption}

\title[HTRU XVII. PSR J1325-6253]{The High Time Resolution Universe Pulsar Survey – XVII. PSR J1325$-$6253, a low eccentricity double neutron star system from an ultra-stripped supernova}

\author[Sengar et al.]{
R. Sengar$^{1,2}$\thanks{E-mail: rsengar@swin.edu.au},
V. Balakrishnan$^{3}$,
S. Stevenson$^{1,2}$,
M. Bailes$^{1,2}$,
E. D. Barr$^{3}$,
N. D. R. Bhat$^{4}$,
M. Burgay$^{5}$,
\newauthor
M. C. i Bernadich$^{3}$,
A. D. Cameron$^{1,2}$,
D. J. Champion$^{3}$,
W. Chen$^{3}$,
C. M. L. Flynn$^{1,2}$,
A. Jameson$^{1,2}$,
\newauthor
S. Johnston$^{6}$,
M. J. Keith$^{7}$,
M. Kramer$^{3,7}$,
V. Morello$^{7}$,
C. Ng$^{8}$,
A. Possenti$^{5,9}$,
B. Stappers$^{7}$,
\newauthor
R. M. Shannon$^{1,2}$,
W. van Straten$^{10}$,
J. Wongphechauxsorn$^{3}$
\\
$^{1}$ Centre for Astrophysics and Supercomputing, Swinburne University of Technology, Mail H39, PO Box 218, VIC 3122, Australia.\\
$^{2}$ARC Center of Excellence for Gravitational Wave Discovery (OzGrav), Swinburne University of Technology, Mail H11, PO Box 218, VIC 3122.\\
$^{3}$Max-Planck Institut f\"ur Radioastronomie, Auf dem H\"ugel 69, D-53121 Bonn, Germany.\\
$^{4}$International Centre for Radio Astronomy Research, Curtin University, Bentley, WA 6102, Australia.\\
$^{5}$ INAF - Osservatorio Astronomico di Cagliari, Via della Scienza 5, I-09047 Selargius (CA), Italy.\\
$^{6}$CSIRO Astronomy $\&$ Space Science, Australia Telescope National Facility, P.O. Box 76, Epping, NSW 1710, Australia.\\
$^{7}$Jodrell Bank Center for Astrophysics, University of Manchester, Alan Turing Building, Oxford Road, Manchester M13 9PL, United Kingdom.\\
$^{8}$Dunlap Institute for Astronomy \& Astrophysics, University of Toronto, 50 St.~George Street, Toronto, ON M5S 3H4, Canada.\\
$^{9}$Universit´a di Cagliari, Dept of Physics, S.P. Monserrato-Sestu Km 0,700 - 09042 Monserrato, Italy.\\
$^{10}$Institute for Radio Astronomy $\&$ Space Research, Auckland University of Technology, Private Bag 92006, Auckland 1142, New Zealand.\\
}

\date{Accepted XXX. Received YYY; in original form ZZZ}

\pubyear{2022}

\begin{document}
\label{firstpage}
\pagerange{\pageref{firstpage}--\pageref{lastpage}}
\maketitle

\begin{abstract}
The observable population of double neutron star (DNS) systems in the Milky Way allow us to understand the nature of supernovae and binary stellar evolution. Until now, all DNS systems in wide orbits ($ P_{\textrm{orb}}>$ 1~day) have been found to have orbital eccentricities, $e > 0.1$. In this paper, we report the discovery of pulsar PSR J1325$-$6253: a DNS system in a 1.81 day orbit with a surprisingly low eccentricity of just $e = 0.064$. Through 1.4 yr of dedicated timing with the Parkes radio telescope  we have been able to measure its rate of advance of periastron, $\dot{\omega}=0.138 \pm 0.002 \, \rm deg \, yr^{-1}$. If this induced $\dot{\omega}$ is solely due to general relativity then the total mass of the system is, $M_{\rm sys} = 2.57 \pm 0.06$\,M$_{\odot}$. Assuming an edge-on orbit the minimum companion mass is constrained to be $M_\mathrm{c,min}>0.98$ M$_{\odot}$ which implies the pulsar mass is $M_\mathrm{p,max}<1.59 $ M$_{\odot}$. 
Its location in the $P$--$\dot{P}$ diagram suggests that, like other DNS systems, PSR J1325$-$6253 is a recycled pulsar and if its mass is similar to the known examples ($>1.3$ M$_\odot$), then
the companion neutron star is probably less than $\sim1.25$\,M$_\odot$ and the system is inclined at about $50^{\circ}$-$60^{\circ}$.
The low eccentricity along with the wide orbit of the system strongly favours a formation scenario involving an ultra-stripped supernova explosion.

\end{abstract}



\section{Introduction}
\label{sec:intro}

Double neutron star (DNS) systems are one of the most important classes of objects used to test and understand numerous astrophysical and fundamental physics phenomena, including general relativity (GR) in the strong-field regime \citep[e.g.,][]{berti_15,ozel_016, kramer_21}. 
They are the systems essentially consisting of two-point masses, whose orbital motion and evolution are well defined by GR.
Mass exchange and tidal effects mean that in close systems, the pre-supernova
orbit was almost certainly circular \citep{Podsiadlowski_92}, and therefore the final
orbital parameters give us an insight into the mass lost in the second supernova explosion, and the magnitude of any asymmetric kicks imparted to the new-born neutron star~\citep[NS; e.g.,][]{sutantyo_78, tauris_98, tauris17}.

In the binary stellar evolution picture, the complex journey of the formation of DNS systems begins with two higher mass stars in a wide binary system~\citep{tauris17,Vigna-Gomez:2018dza}. 
The higher-mass star evolves first and undergoes a supernova explosion which results in the formation of an NS in a highly bound eccentric orbit with a massive star companion such as PSR B1259--63 \citep{johnstn_92, kaspi_94}. Subsequently, the system enters the high-mass X-ray binary phase, where the companion is close to filling its Roche-lobe, and starts transferring matter onto the NS \citep{vanden_019}. This phase of mass-transfer is dynamically unstable as mass is being transferred from the more massive companion to the less massive NS, and results in a runaway mass transfer process and hence the formation of a common-envelope \citep[CE;][]{Paczynski:1976IAUS} phase. During the CE phase, upon expansion, the companion star engulfs the NS and the drag forces cause the binary to spiral closer. Under such circumstances, the first born NS may be spun-up to a spin period of a few tens of milliseconds through accretion~\citep[][]{Chattopadhyay:2019xye}. Additional mass transfer after the CE phase can further spin up the pulsar, and reduce its magnetic field strength in the process~\citep{tauris17}. These phases are expected to circularize the orbit of the binary \citep[e.g.,][]{belczynski_08}. Once the companion reaches the end of its evolution, it also undergoes a supernova explosion and forms another NS. The imparted kicks and mass-loss to the system during the final supernova explosion may disrupt the system and result in a pair of isolated NSs i.e. a young and a recycled pulsar. However, in at least some systems the binary may survive both supernovae (SNe), and a bound pair of NSs in an eccentric orbit is formed where the second born NS is a young/normal pulsar. The detectability of the system depends upon the relative luminosities of the two pulsars, their beaming fraction, beam orientation, and age. Many young pulsars perish after just a few Myr, whereas recycled pulsars can live for Gyrs. As a result, the vast majority of the DNS systems only have the recycled pulsar visible. The sole exception was the double pulsar PSR J0737$-$3039A/B, but due to precession the B pulsar is no longer visible \citep{burgay_03,lyne_04,kramer_21}.\par

\begin{table*}
\centering
\caption{Binary parameters of the known Galactic DNS systems in ascending order of their right ascension (RA). The solid black line in between divides systems with well measured masses and constrained masses based on their $\dot{\omega}$ measurements. The two DNS systems which reside in globular clusters (PSRs J1807$-$2459 and B2127$+$11C) are not included here. Two potential DNS candidates (PSRs J1755$-$2550a and PSR J1753$-$2240) are excluded from the table due to the absence of an $\dot{\omega}$ measurement from their timing.}
\label{tab:DNS_systems}
\begin{tabular}{lllllllll}
\hline
\hline
Pulsar & Spin period & $P_{\rm orb}$ & $e$ & $\dot{\omega}$ & $f(M_{\rm p},M_{\rm c})$ & $M_{\rm c}$ & $M_{\rm p}$ & $M_{\rm sys}$ \\
       &(ms)         & (days)        &     &($\rm ^{o}yr^{-1}$)& ($\rm M_{\odot}$) & ($\rm M_{\odot}$) & ($\rm M_{\odot}$) & ($\rm M_{\odot}$) \\
\hline
J0453$+$1559$^{1 \rm }$     & 45.782$^{\rm b}$      &4.072    &0.11251844  &0.03793      & 0.196012     &1.174(4)      &1.559(5)      &2.733(4)\\
J0509$+$3801$^{2}$          & 76.541$^{\rm b}$      &0.379    &0.586400    &3.031        &   0.064255   &1.46(8)       &1.34(8)       &2.805(3)  \\
J0737$-$3039A/B$^{3}$       & 22.699$^{\rm b}$      &0.102    &0.08777023  &16.899323    & 0.290987     &1.248868(13)  &1.338185(12)  &2.587052(9)\\
J1518$+$4904$^{4}$          & 40.935$^{\rm b}$      &8.634    &0.24948451  &0.0113725    &0.115988      &1.31(8)       &1.41(8)       &2.7183(7)    \\
B1534$+$12$^{5}$            &37.904$^{\rm b}$       &0.420    &0.27367752  &1.7557950    &0.314630      &1.3455(2)     &1.3330(2)     & 2.678463(4)\\
J1756$-$2251$^{6}$          &28.461$^{\rm b}$       &0.319    &0.1805694   &2.5825       &0.220109      &1.230(7)      &1.341(7)      & 2.56999(6) \\
J1757$-$1854$^{7}$          &21.497$^{\rm b}$       &0.183    &0.605816878 &10.365001    &0.357188      &1.3946(9)     &1.3384(9)     &2.73295(9)  \\ 
J1829$+$2456$^{8}$          &41.009$^{\rm b}$       &1.176    &0.1391412   &0.2919       &0.294377      &1.299(4)      &1.306(4)      &2.60551(19)  \\
J1906$+$0746$^{9a}$         &144.073$^{\rm c}$      &0.165    &0.0853028   &7.5841       &0.111566      &1.322(11)     &1.291(11)     &2.6134(3)  \\
J1913$+$1102$^{10}$         &27.285$^{\rm b}$       &0.206    &0.089&5.63  &0.136344     &1.27(3)       &1.62(3)       &2.8887(6)   \\
B1913$+$16$^{11}$           &59.031$^{\rm b}$       &0.3228   &0.6171340   &4.226585     &0.132167      &1.3886(2)     &1.4398(2)     &2.828378(7)  \\
\hline
J1411$+$2551$^{12}$         & 62.453$^{\rm b}$      &2.615    &0.1699308   &0.0768       &0.122390      &$>$0.92       &$<$1.64       &2.538(22)\\
J1759$+$5036$^{13}$         &176.016$^{\rm b}$      &2.042    &0.30827     &0.127        &0.081768      &$>$0.7006     &$<1.8$        &2.62(3)\\
J1811$-$1736$^{14}$         &104.182$^{\rm b}$      &18.779   &0.828011   &0.0090        &0.128121      &$>$0.91       & $<$1.75      &2.57(10)  \\
J1930$-$1852$^{15}$         &185.520$^{\rm b}$      &45.060   &0.39886340   &0.00078     &0.346908      &$>$1.30       &$<$1.32       &2.59(4)  \\
J1946$+$2052$^{16}$         &16.960$^{\rm b}$       &0.078    &0.063848     &25.6        &0.268073      &$>$1.18       &$<$1.32       &2.50(4)  \\
\textbf{J1325$-$6253}$^{\rm This \, work}$          & $28.968^{\rm b}$     &1.815        &0.06400       &0.138         &0.141516      & $>$0.98 & $<1.59$ & 2.57(6)\\
\hline
\end{tabular}
\\
\begin{flushleft}
\textbf{References:}
(1) \citet{martinez015}, 
(2) \citet{lynch018},
(3) \citet{kramer_21},
(4) \citet{janssen08},
(5) \citet{fonseca014},
(6) \citet{ferdman014},
(7) \citet{cameronJ1757018},
(8) \citet{champion_05,Haniewicz_21},
(9) \citet{leeuwen015},
(10) \citet{ferdman_018,ferdman2020},
(11) \citet{weisberg010},
(12) \citet{martinez017},
(13) \citet{agazie021},
(14) \citet{corongiu07},
(15) \citet{swiggum015},
(16) \citet{stovall018}.\\
{\footnotesize ${^{\rm a}}$ Unconfirmed DNS system: The companion could be a white dwarf. The values of $M_{\rm c}$ and Mc for PSR J1518$+$4904 are taken from section 8.3 in \citet{tauris17}}\\
{\footnotesize ${^{\rm b}}$ Recycled pulsars }, {\footnotesize ${^{\rm c}}$ Young pulsars }

\end{flushleft}
\end{table*}

The first such DNS system, PSR B1913+16, was discovered in 1974 \citep{hulseandtaylor75}. 
Precise timing of this system ultimately provided the first indirect evidence of gravitational wave (GW) emission due to its orbital damping~\citep{Taylor:1982ApJ,weisberg010}. This binary pulsar provided the impetus to build the Laser Interferometer Gravitational-Wave Observatory (LIGO) for the direct detection of GWs as it provided proof of the existence of detectable sources of NS mergers in our Universe.
Since then 20 additional DNS systems (including candidate DNS systems where the companion's identity is unconfirmed) have been discovered in sensitive large-scale pulsar surveys, like the Parkes-multibeam pulsar survey \citep[PMPS;][]{pmps01}, the Arecibo L-Band Feed Array pulsar survey \citep[PALFA;][]{alfa_06,lazarus15}, the Parkes High-Latitude pulsar survey \citep[PHPS;][]{phps_06}, the High Time Resolution Universe Survey of the southern Galactic plane \citep[HTRU;][]{keith10} and The Green Bank North Celestial Cap survey \citep[GBNCC;][]{gbncc_14}. We list some properties of the known and candidate Galactic DNS systems in Table~\ref{tab:DNS_systems}.
Many of these DNS are in compact orbits that will merge approximately within a Hubble time, and are highly relativistic in nature. 
Typically, several Post-Keplerian parameters can be measured within a few years through precision timing, allowing for tests of theories of GR and other theories of gravity \citep[e.g.][]{kramer_21}. Using precision timing, the inferred merger rate and precise NS mass measurements provided by these systems were instrumental in predicting the waveforms and detection rates expected for the Advanced LIGO and Virgo gravitational-wave detectors \citep{abbott17,LIGOScientific:2020aai} in our local universe i.e. within the detection horizon of the instruments. \par
Indeed, in 2017 August, the LIGO--Virgo network witnessed the spectacular merger of two NSs \citep[GW170817;][]{abbott17}, accompanied by the detection of an unusually low-luminosity short-duration gamma-ray burst~\citep{LIGOScientific:2017zic}.
The estimated masses of the NSs were consistent with those seen in our own Galaxy \citep[][]{abbott17}.
In their third major observing run, a second DNS merger was observed, with an estimated total system mass of $3.4$\,M$_\odot$, significantly higher than any DNS known in our Galaxy~\citep{LIGOScientific:2020aai,Galaudage:2020zst}.
Heavier mass systems decay faster than lighter ones, are more luminous in GWs, and have a shorter radio pulsar lifetime \citep{chaurasia05}. 
All of these selection effects may bias the relative populations of BNS mergers and Galactic binary radio DNS systems.

DNS systems in wide orbits ($P_{\rm orb} \gtrsim 1$\,d) will usually not merge within a Hubble time (unless they have extremely high eccentricities), and their pulsar ages are such that their orbital parameters i.e. $P_{\rm orb}$ and $e$ have not been greatly affected by the GW emission to any significance. 
Their orbits therefore closely resemble that of the immediate post-SN state and can be used to constrain the mass-loss and kicks NSs receive at birth in such systems. 
In this paper we present the discovery and timing analysis of the binary pulsar PSR J1325$-$6253 which was discovered in a reprocessing of the High Time Resolution Universe South Low Latitude (HTRU-S LowLat) pulsar survey~\citep{keith10, cherry15, cameron_020}. 
PSR J1325$-$6253 is the second DNS system discovered in the survey after PSR J1757$-$1854, which remains the most accelerated known pulsar in a Galactic DNS system \citep{cameronJ1757018}. In Section~\ref{sec:Observations and Data reduction}, we first describe the discovery of PSR J1325$-$6253 in the HTRU-S LowLat survey and then timing analysis of the observations taken with the ultra-wide-band receiver (UWL) of the Parkes-radio telescope (also known as \texttt{Murriyang}). In Section~\ref{sec:properties}, we summarise the main properties of PSR J1325$-$6253 and its orbit. We describe our search for (and lack of detection of) the companion in Section~\ref{sec:companion_search}. We then discuss the implications of the discovery of PSR J1325$-$6253 in Section~\ref{sec:discussion}. Finally, we draw our conclusions in Section~\ref{sec:conclusion}.

\section{Observations and Data analysis} \label{sec:Observations and Data reduction}

\subsection{Discovery of PSR J1325$-$6253 in HTRU-S LowLat}
\label{subsec:discovery_in_HTRU_Lowlat}

PSR J1325$-$6253 was discovered in a GPU accelerated reprocessing of the HTRU-S LowLat survey (Sengar et al. in preparation) which employs the time-domain resampling method \citep{johnstn_92} to search for binary pulsars. 
In this reprocessing, the 72 minutes of the full-time resolution data were searched coherently with a series of constant acceleration trials between $\pm \rm  50 ~ms^{-2}$. PSR J1325$-$6253 was detected in beam 10 of an observation starting at UTC 2011-12-10-16:54:46 using the multibeam receiver \citep[MB;][]{multibeam96} of the Parkes 64-m Radio Telescope with an acceleration $a=-3.5$\,m\,s$^{-2}$, and possessed a fast Fourier Transform (FFT) signal-to-noise ratio (S/N) of 9.7 in the fourth harmonic fold. 
The original filterbank data were then folded at the detected period of 28.96\,ms, and trial dispersion measure of 303\,$\rm pc \, cm^{-3}$ and a search for the optimal period and acceleration was conducted using the \texttt{psrchive} routine \texttt{pdmp}~\citep{hotan_04}. 
After folding, an optimized S/N of 14.3 was obtained, well above the false alarm threshold, $\rm S/N_{thresh} = 9$ of our search.
The non-detection of PSR J1325$-$6253 in previous processing of the survey conducted by \cite{cherry15} and \cite{cameron_020} was due to the segmented nature of their pipeline. In those searches, the 72-min full-length observations were only searched for accelerations, $|a| \leq 1 \, \rm m\,s^{-2}$ and higher acceleration searches were only performed on segmented observations where the sensitivity of a pulsar signal is degraded by a factor of $\sqrt{N}$, where $N$ in the number of segments of the full integration observation.
In the two-segment search, the pulsar's S/N in the FFT search would have only been $\sim 7$ which is below $\rm S/N_{thresh} = 8$ in previous searches of the survey.

\subsection{Detection of PSR J1325$-$6253 in PMPS data} 
\label{subsec:detection_in_pmps}

A major sky fraction of the PMPS survey covers the entire HTRU-S LowLat, therefore a search of the closest (if any) PMPS observation is  warranted to determine if PSR J1325$-$6253 is detectable in the earlier survey. Three such observations were found, each with a different epoch but with the same coordinates as one another. These observations are 4.6-arcmin away from the beam centre of the discovery observation of PSR J1325-6253 in the HTRU-S LowLat survey. Although both 
surveys used the same MB receiver with almost same bandwidth, but they were different in digitisation, frequency, and time-resolution i.e. PMPS data are 1-bit digitized, has eight times more channel bandwidth (3MHz) and four times coarser time-resolution (250 $\mu$s). 
More importantly, it has half the integration length (35 mins) as compare to 72-min HTRU-S LowLat observations. All these factors make PMPS less sensitive than the HTRU-S LowLat survey. PSR J1325$-$6253 was discovered with S/N=14.3 in the HTRU-S LowLat, if the PMPS observations had the same coordinates as its discovery observation then, by scaling for the difference in integration times between the two surveys, the pulsar should have been detected with S/N$=14.3\sqrt{35/72}\sim 10$. However, for the MB receiver, an offset of 4.6-arcmin results in a degradation in sensitivity of 27\%, as the response pattern of the MB receiver is approximately Gaussian. Any offset from the discovery position of the pulsars would decrease the S/N by a factor of $e^{\theta^2/-2 \sigma^2}$\footnote{where $\theta$ is the beam offset in degrees and $\sigma$ is related to the MB receiver's beam FWHM by $\sigma=\dfrac{\textrm{FWHM}}{2\sqrt{2\textrm{ln}2}}$.}, i.e. PSR J1325$-$1325 would be detected with S/N$\sim$7.5 in PMPS. We first performed an acceleration search using the time-domain resampling method \citep[e.g.,][]{middleditch84,johnston92a} on those three PMPS observations and the pulsar was detected with FFT S/N$\sim$6 and acceleration, $a\sim-3.0$ ms$^{-2}$. Also, upon direct folding the observations with the ephemeris given in Table \ref{table:ephemeris_table} the pulsar signal was recovered. From both methods we got the expected S/N of $\sim$ 7.5--8. The reason why this pulsar was missed in initial processing of the PMPS data is mainly due to its low statistical significance. First, when conducting acceleration trials and searching 1000s of DM trials the $\rm S/N_{thresh}$ is much greater than 6. This results in the vast majority of 6$\sigma$ candidates being spurious. Secondly, at the DM of the pulsar (303 $\rm pc \, cm^{-3}$) the smearing of the profile is $\sim$11\% of the pulse width using the PMPS analogue filterbanks, the 1-bit PMPS filterbank system is about 15\% less sensitive than the HTRU digital filterbank system.

\subsection{Timing analysis} \label{sec:discovery_and_timing}

As discussed in section \ref{subsec:discovery_in_HTRU_Lowlat}, the presence of acceleration indicated that PSR J1325$-$6253 was in a binary system. We confirmed the pulsar with the Parkes MB receiver on 2019 September 26, when it was again detected with similar acceleration and a significantly different spin period, confirming its binary nature. 
After the decommissioning of MB receiver from 2020 October, the initial follow-up observations were conducted using the Parkes UWL receiver \citep{uwl} in search mode with 128-$\mu s$ time-resolution and 128 MHz channel bandwidth. 
These initial observations allowed us to obtain its rough orbital parameters; in particular, its orbital period, $P_{\textrm{orb}}\sim1.81$ d and eccentricity, $e\sim0.064$ using the Fortran computer program \texttt{fitorbit}\footnote{\url{https://github.com/vivekvenkris/fitorbit}}. 
At this stage however, it was impossible to obtain a phase connected timing solution as it is normally difficult to separate the pulsar position from other parameters until a year of data is obtained without an interferometric position. To obtain the latter we observed the pulsar with the TRAPUM\footnote{\url{www.trapum.org}} backend \citep[TRAnsients and PUlsars with MeerKAT,][]{trapum_16} on the newly-commissioned MeerKAT telescope by forming many tied beams near the position of the pulsar on 2020 October 21, and folding at the period of the pulsar using the TRAPUM Seekat\footnote{\url{https://github.com/BezuidenhoutMC/SeeKAT}} system.
The best TRAPUM position was 4.18-arcmin away from the coordinates of the HTRU-S LowLat discovery beam and made subsequent timing observations with the Parkes telescope more efficient. It also enabled a quick phase-connected timing solution to be obtained from the existing data and helped to obtain an approximate orbital solution for this pulsar, however the measurement of the rate of advance of periastron, $\dot{\omega}$, was not significant enough to constrain the total system mass.
From 2021 April, we initiated a dedicated observational campaign with the Parkes radio telescope for this pulsar with a monthly cadence, aiming to observe the full orbit at multiple epochs. This quickly removed the covariances between the various orbital parameters and the pulsar position. In order to cover the entire phase of the orbit, during each campaign we observed it for $\sim$22 hr over a time span of two consecutive days. All observations were recorded using coherent dedispersion with four polarisations i.e. Stokes $I$, $Q$, $U$, and $V$ spectra along with a noise diode observation before each set of observations. The time-resolution for each observation was 128\,$\mu$s, with a bandwidth of 1\,MHz and 3328-frequency channels. Prior to the timing analysis, we cleaned the folded archive file first using \texttt{clfd}\footnote{\url{https://github.com/v-morello/clfd}} \citep{morello18} and then used \texttt{paz} of the \texttt{PSRCHIVE} package \citep{hotan_04}. As a next step of radio-frequency-interference (RFI) excision we manually inspected each observation and deleted the bad channels using \texttt{pazi} of the \texttt{PSRCHIVE} package. The same RFI excision procedure was applied to noise diode observations and observations were polarization and flux calibrated. \par

\begin{figure}
\centering
\includegraphics[width=0.5\textwidth]{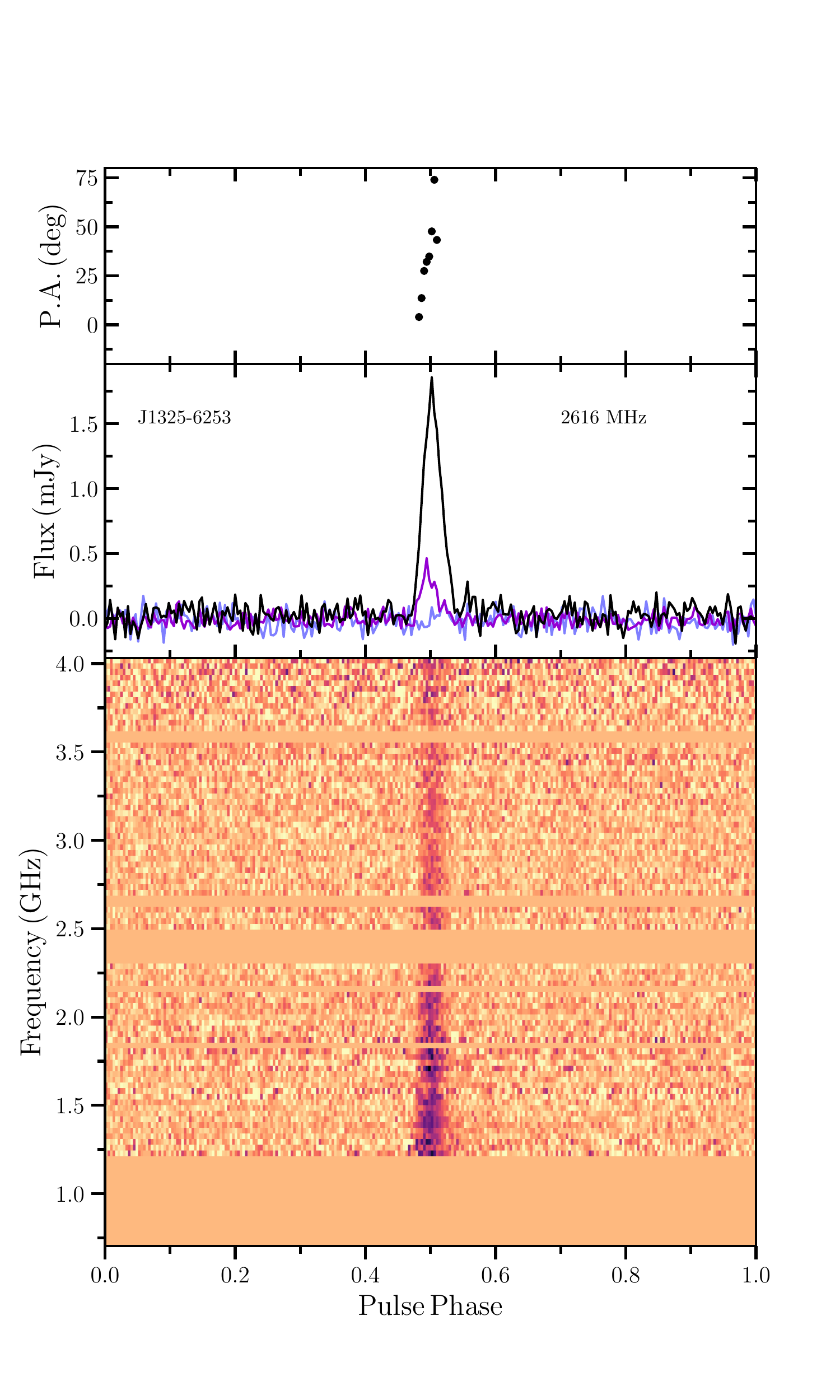}
\caption{The bottom panel shows the frequency-phase plot of PSR J1325$-$6253 across a large portion of UWL band (1200--4000\,MHz), obtained by a 7.4-h observation. The intensity values in this plot are saturated at the fifth percentile. The flux and polarization calibrated pulse profile is shown in middle panel where the total intensity is represented by the black line. The polarization profile was obtained by dedispersing and defaradaying the polarization profiles in each frequency channel before summing at the RM of --144 $\rm rad \, m^{-2}$. Red and blue lines represent linear and circular polarizations respectively. The top panel represents position angle (PA) in the leading edge of the pulse profile derived from the linearly polarized flux.}
\label{fig:pulse_profile}
\end{figure}

\begin{figure*}
\centering
\includegraphics[width=0.9\textwidth]{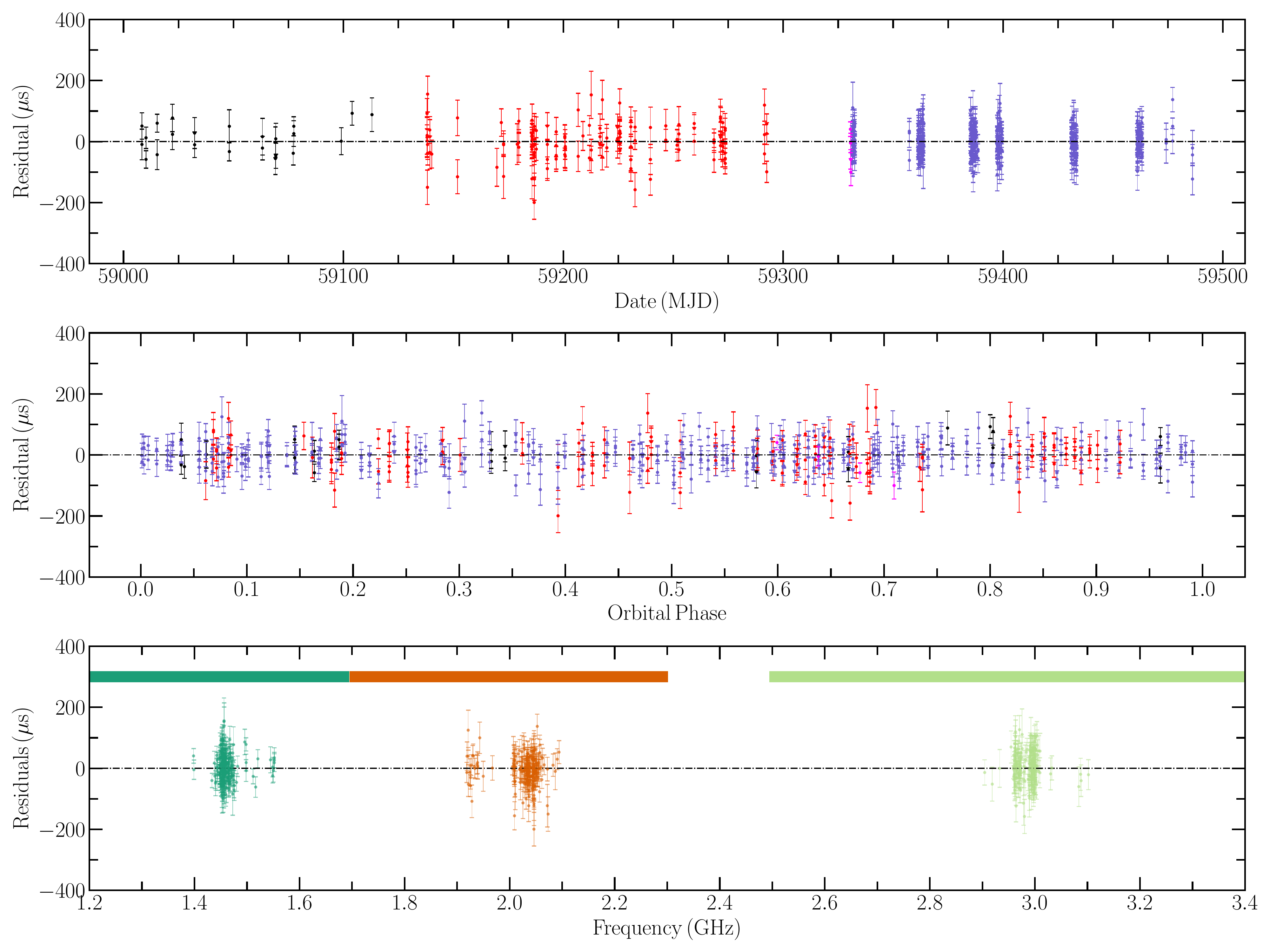}
\caption{The top panel shows the timing solution for PSR J1325$-$6253, where post-fit timing residuals are plotted as a function of MJD. The reduced $\chi^{2}$ ($\chi_{\rm red}$) value of 1 was obtained by applying a multiplicative error factor (EFAC) of 1.1. Data points in black correspond to the observations taken in incoherent search mode, red represents uncalibrated coherent fold mode data, blue is the polarisation and flux calibrated data in coherent fold mode and pink represents data taken with the DFB4 backend. The middle panel shows residuals vs orbital phase. The bottom plot shows residuals vs central frequency at which TOAs were generated. The TOAs centred near 1.45 GHz, 2.05 GHz and 3 GHz are shown in green, orange and light green colors. The thick border lines with same colors correspond to the bandwidth  used to create those TOAs. The spread in frequency of each group of TOAs is due to RFI excision of bad channels which changed the effective mean frequencies of the TOAs. The frequency band from 2300$-$2500\,MHz is excluded from the analysis as this part of the band is badly RFI-affected.  }
\label{fig:toas}
\end{figure*}

From full-band observations (704--4032\,MHz), PSR J1325$-$6253 is visible from 800--3400\,MHz. However, we restricted our analysis to the frequency band between 1200 and 3400 MHz, as for most observations the lowest portion of the band is typically contaminated with RFI. Additionally, the highly scattered profile of the pulsar at these lower frequencies makes it difficult to derive quality times-of-arrival (TOAs). The integrated pulse profile of PSR J1325$-$6253 between 1200 and 3400 MHz is shown in Figure \ref{fig:pulse_profile} which has been created by adding multiple observations with total integration time of 7.4 hrs taken during one of the full orbit campaigns. The observation has been polarization and flux calibrated. From Figure \ref{fig:pulse_profile}, it is evident that the pulsar is 25 per cent linearly polarized and has very little circular polarization. The observations were typically $\sim$ 1-h in integration time. 
We summed these observations in the time domain, and divided them into three separate frequency bands with bandwidths 1200--1700, 1700--2300 and 2500--3400\,MHz centered at 1450, 2000, and 2950\,MHz respectively, to obtain three mean pulse profiles per observation and pulse profiles with low significance i.e. $\rm S/N<8.5$ were discarded before the timing analysis. In total, out of the sub-banded observations, 548 remained for timing analysis. Corresponding to each sub-band, we created three standard template profiles from the same observation shown in Figure \ref{fig:pulse_profile}. 
These templates were then used to compare with 548 real pulse profiles of the pulsar to obtain the TOAs using \texttt{PSRCHIVE} routine \texttt{pat}. 
Finally, \texttt{TEMPO2} was employed using the theory independent DD binary model \citep{damour_86} to describe the orbital motion and obtain a complete phase-connected solution. The timing residuals using this model are shown in Figure~\ref{fig:toas}. The complete ephemeris obtained from \texttt{TEMPO2} is given in Table \ref{table:ephemeris_table}. 

\begin{table}
\caption{Timing Parameters of PSR J1325$-$6253}

\centering
\resizebox{\columnwidth}{!}{%
\begin{tabular}{ll}
\hline
PSR & J1325$-$6253 \\ \hline
Fitting program & TEMPO2\\
Time units  & TCB \\
Solar system ephemeris & DE405 \\
Span of timing data (MJD/days)  &  58981-59505/524  \\
Number of TOAs & 548   \\
rms residuals ($\mu \rm s$) & 42\\
R.A., $\alpha$(J2000)  & 13:25:04.8907(2) \\
Decl., $\delta$(J2000)  &    --62:53:39.594(2) \\
Galactic longitude, $l$($^{\circ}$)  &  306.753   \\
Galactic latitude,  $b$($^{\circ}$) &     --0.272 \\
Spin period, $P$ (s)   &  0.02896859323602(4) \\
Period derivative, $\dot{P} \, (\rm 10^{-20}ss^{-1})$ & 4.80(14)\\
Dispersion measure ($\rm pc \, cm^{-3}$) &  303.331(3)\\
\hline
Binary Parameters &  \\ 
\hline
Orbital model  &   DD    \\
Orbital period, $P_{\rm orb}$ (days) &   1.81559019(7)\\
Orbital eccentricity, $e$   &   0.0640091(7)   \\
Projected semimajor axis, $x$(lt-s) & 7.573914(3)   \\
Epoch of periastron, $T_{0}$ (MJD) &   58979.06160  \\
Longitude of periastron, $\omega (^{\circ})$ & 190.233(3) \\
Rate of advance of periastron, $\dot{\omega}(\rm ^{o}yr^{-1})$    &     0.138(2) \\
\hline
Derived Parameters & \\ 
\hline
Mass function, $f(M_{\rm p}, M_{\rm c})$ ($\rm M_{\odot}$)   &  0.1415168(1)    \\
Total system mass, $M_{\rm sys} \, (\rm M_{\odot})$  &  2.57(6)   \\
Maximum Pulsar mass, $M_{\rm p, max}\, (\rm M_{\odot})$&  $<$1.59  \\
Minimum companion mass, $M_{\rm c, min}\, (\rm M_{\odot})$  & $>$0.98    \\
DM derived distance, $D_{\rm NE2001} \rm (kpc)$ &  4.4    \\
DM derived distance, $D_{\rm YMW16} \rm (kpc)$  &  5.4   \\
Surface magnetic field strength, $B_{0}(10^{9} \rm G)$             & 1.18(17)     \\ 
Spin down luminosity, $L(10^{31} \rm erg \, s^{-1})$                & 7.80(23) \\
Characteristic age, $\tau_{c}$ (Gyr)                               &  9.5(3)   \\   
Flux density at 1.4 GHz, $S_{1.4}$ (mJy)                            & 0.13    \\
Flux density at 3.8 GHz, $S_{3.8}$ (mJy)                            & 0.035    \\
Spectral index, $\alpha$                                          &  $-$1.2(1)   \\
Rotation measure, RM ($\rm rad \, m^{-2}$)                      & --144(3) \\

\hline
\end{tabular}}
\label{table:ephemeris_table}
\end{table}

\section{Properties of PSR J1325$-$6253} 
\label{sec:properties}

\begin{figure}
\centering
\includegraphics[width=0.45\textwidth]{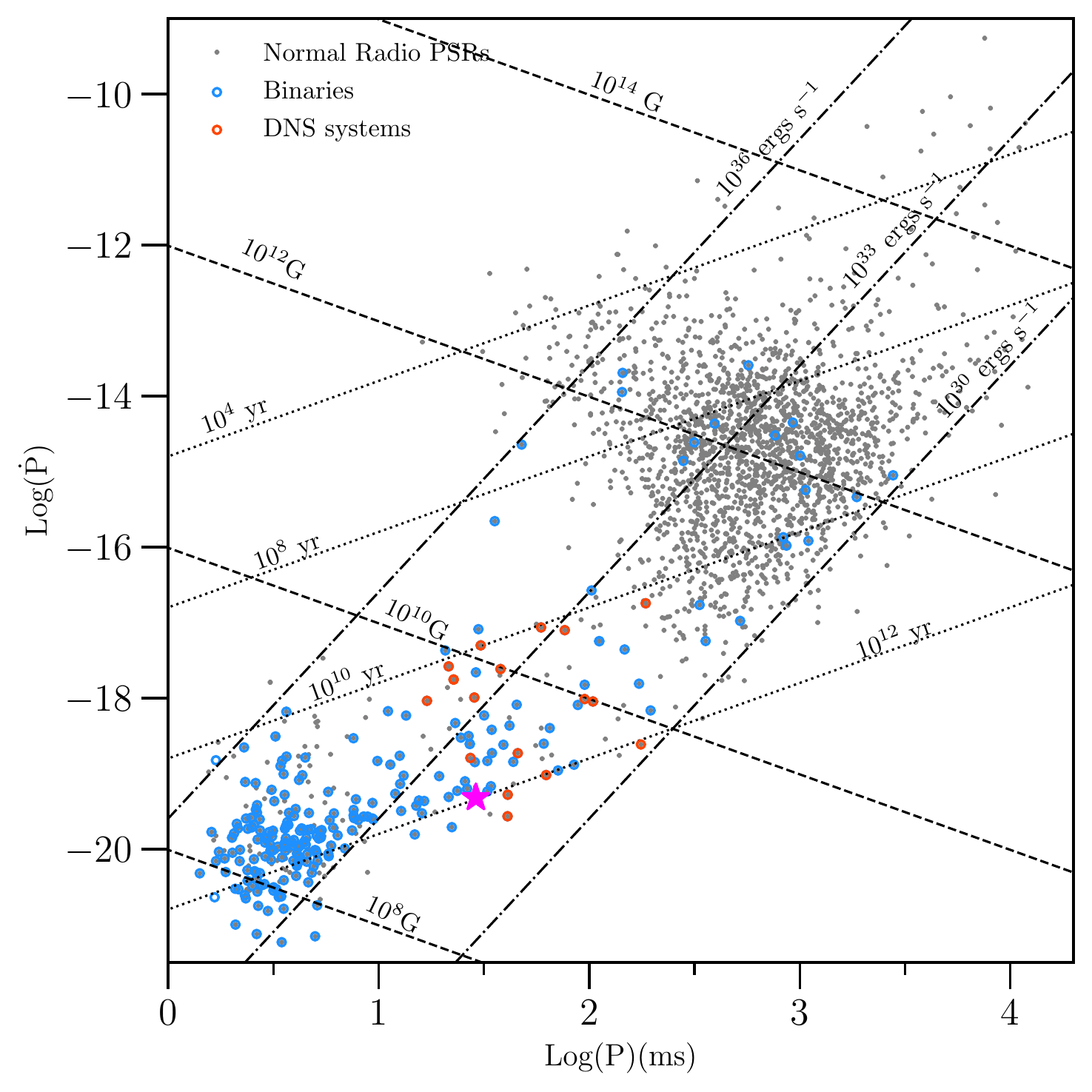}
\caption{ $P$--$\dot{P}$ diagram of all known pulsars including known DNS systems (in red), other binaries (in blue) and normal pulsars (in black). The measured value of $P$ and $\dot{P}$ are taken from the ATNF pulsar catalogue \citep{psrcat05a} version 1.6. It is clearly evident that PSR J1325$-$6253 (in magenta color) belongs to the population of other recycled DNS systems.}
\label{fig:ppdot}
\end{figure}

Based on the precisely measured spin period, $P$ and period derivative $\dot{P}$ for J1325$-$6253 (provided in Table \ref{table:ephemeris_table}), we show the location of PSR J1325$-$6253 in the $P$--$\dot{P}$ diagram in Figure \ref{fig:ppdot}.
PSR J1325$-$6253 falls in the region occupied by other known recycled DNS systems and far from that of the unrecycled pulsars. We therefore assert that PSR J1325-6253 is the recycled NS in this binary system, as opposed to the young, second-formed companion NS. Using $P$ and $\dot{P}$ we measured its minimum characteristic age, $\tau_{\rm c}$ and maximum surface magnetic field strength, $B_{\rm surf}$ to be 9.5\,Gyr and $1.1 \times 10^{9}$\,G respectively. These values are also in agreement with the values of other recycled binary pulsars with massive companions~\citep[e.g.,][]{tauris17}.
The DM-derived distance for PSR J1325$-$6253 is 4.4 or 5.4\,kpc, based on NE2001 \citep{ne2001} and YMW16 \citep{ymw16} electron density models, respectively.  \par

Although we cannot yet determine the proper motion, and hence the transverse velocity of the system, at its likely distance of 4-5 kpc, we conclude that the observed period is unlikely to be greatly affected by the Shklovskii effect~\citep{Shklovskii:1970SvA}. 
Even if the pulsar had a transverse velocity of 100\,km\,s$^{-1}$, and was at a distance of 5\,kpc, its observed $\dot P$ would only be contaminated by about 15\%. 
As we argue later on in Section~\ref{sec:discussion}, it is unlikely that this binary received a large kick in the second supernova explosion and is unlikely to
have a large transverse velocity.

The calibrated flux density of J1325$-$6253 across the 1200-4000\,MHz UWL-band allowed us to measure its spectral index. 
We divided the 7.4-h flux-calibrated observation shown in Figure \ref{fig:pulse_profile} into seven-different sub-bands (with 400-MHz bandwidth each), and for each band we formed a standard template which is cross-correlated with the observed profile using the \texttt{PSRCHIVE'S} program \texttt{psrflux} and the average flux density was obtained for each band. 
At 1.4\,GHz we obtained a flux density ($S_{1400}$) of 0.13\,mJy and at 3.8\,GHz, $S_{3800}$ is 0.035\,mJy. 
A power-law fit of the form $S_{\nu}\sim A\nu^{\alpha}$ to the flux densities at seven frequencies results in a spectral index, $\alpha = -1.2 \pm 0.1$, not dissimilar to many pulsars \cite{jankowski018}. We also measure the rotation measure of the pulsar using \texttt{psrchive} program \texttt{rmfit} and found it to be --144 $\pm$ 3 $\rm rad \, m^{-2}$.

\subsection{Nature of the binary system}
\label{sec:system_nature}

For DNS systems in tight orbits, multiple post-Keplerian (PK) parameters can typically be measured, and the mass of the pulsar and its companion can be measured precisely (assuming the correctness of GR). However, for binary systems with orbital periods of roughly a day or longer, only $\dot{\omega}$ and Shapiro delay (if the system is edge-on) can be measured. Measuring the remaining PK parameters is a challenging task and may require decades of precision timing. Using the relation from \cite{peters64}, the merger time-scale, $\tau_{\rm GW}$ for PSR J1325$-$6253 is 212 Gyr which suggests like other DNS systems in wide binaries, it will also not merge within a Hubble time and will not effect the DNS merger rates predicted by LIGO--Virgo detector network \citep{LIGOScientific:2020aai}. The well-measured orbital eccentricity, $e = 0.064$ of PSR J1325$-$6253 allows the rate of advance of periastron, $\dot{\omega}$ to be measured in this system. From the dedicated timing of PSR J1325$-$6253, we were able to measure $\dot{\omega}$ to be $0.138^{\circ}\pm 0.002^{\circ} \, \rm yr^{-1}$ which has a high significance of $\sim$70~$\sigma$. If we assume that the induced $\dot{\omega}$ is solely due to the relativistic effects (as in DNS systems) then the total system mass can be calculated using relation \citep{taylor_82},

\begin{equation}
\label{eq:m_sys}
M_\mathrm{sys} = \dfrac{1}{T_{\odot}} \bigg(\dfrac{P_{\textrm{orb}}}{2\pi}\bigg)^{5/2}[\dot{\omega}(1-e^{2})/3]^{3/2} ,
\end{equation}
where $M_\mathrm{sys} = m_{\rm p} + m_{\rm c}$ is the total system mass, $m_{\rm p}$ and $m_{\rm c}$ refer to the pulsar and companion masses. 
Using the above equation we found $M_\mathrm{sys}=2.57 \pm 0.06 \rm \, M_{\odot}$ which is typical for Galactic DNS systems \citep[e.g.,][]{Farrow:2019xnc} and is comparable to the DNS system PSR J1811$-$1736, with $M_\mathrm{sys} = 2.57(10) \, \rm M_{\odot}$ \citep{corongiu07}.
In order to constrain the mass of the unseen binary component, we can use a quantity called the ``mass function" which relates the orbital parameters of a binary system i.e. $P_{\rm orb}$ and the projected semi-major axis, $x = a \sin(i)$, to the masses of both components. 
For PSR J1325$-$6253, we obtain the mass function using

\begin{align}
\label{eq:mass_function}
f(M_{\rm p},M_{\rm c}) = \dfrac{4 \pi^{2} x^{3}}{P_{\textrm{orb}}^2 T_{\odot}} = \dfrac{(M_{\rm c} {\sin}i)^3}{M_{\textrm{sys}}^2} = 0.1415 \, \textrm{M}_{\odot},
\end{align}
where $T_{\odot}=G\textrm{M}_{\odot}/c^{3}=4.925490947\mu s$ is used to express masses in solar units, $i$ is the angle between the orbital plane and plane of the sky as observed from Earth. 
To obtain the minimum companion mass $M_{\rm c}$ we can assume an edge-on orbit i.e., $\sin(i) = 1$ and using the $M_\mathrm{sys}$ the cubic polynomial $M_{\rm c}^{3}-f(M_{\rm p},M_{\rm c})M_{\textrm{sys}}^{2}>0$ can be solved. 
Given the value of $M_{\textrm{sys}}$ and $f(M_{\rm p},M_{\rm c})$ from Equations \ref{eq:m_sys} and \ref{eq:mass_function}, we find $M_{\rm c}>0.98 \, \textrm{M}_{\odot}$ which implies $M_{\rm p}<1.59 \, \textrm{M}_{\odot}$.

\begin{figure*}
\centering
\includegraphics[width=0.8\textwidth]{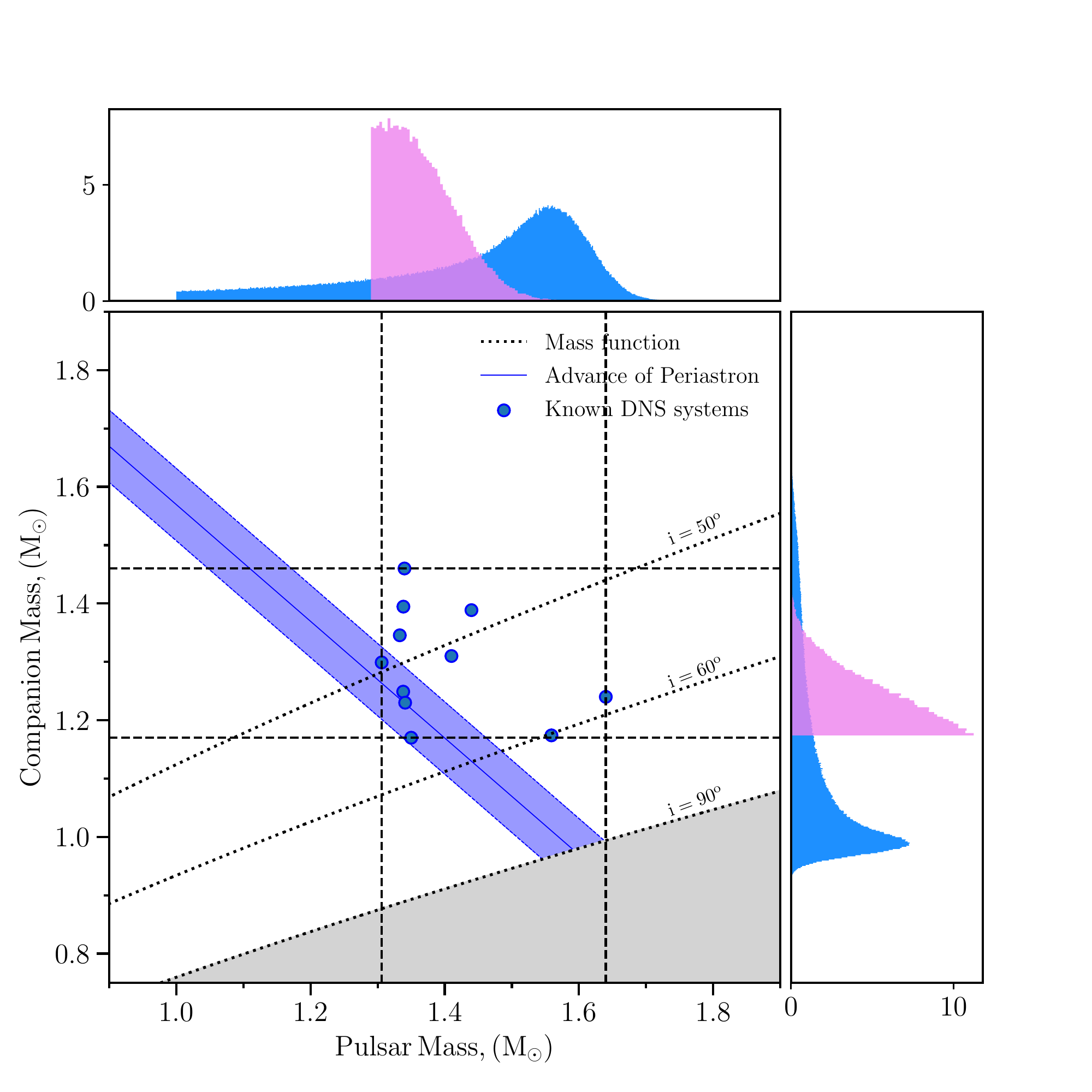}
\caption{Mass--mass diagram for PSR J1325$-$6253 showing constraints on the pulsar mass and the companion mass. The panels on the top and to the right show the probability distribution functions for the pulsar and companion masses assuming an isotropic inclination distribution (in light blue) and distribution using the astrophysical priors (pink). The grey region in the main panel is excluded due to the constraint on the binary mass function corresponding to the maximum possible orbital inclination angle i.e., $\sin(i) = 1$. Other black dotted lines represent companion mass limits for various orbital inclinations. The solid blue line represent the measured $\dot{\omega}$ and the dotted blue lines represent $\dot{\omega}$ within 1$\sigma$. The vertical and horizontal dashed black lines correspond to the minimum and maximum pulsar and companion masses among the known DNS systems for which both masses are well measured. The well-measured masses for individual DNS systems are shown in blue points.}

\label{fig:mass_mass_diagram}
\end{figure*}

The constraints on these masses are shown in the mass--mass diagram for J1325$-$6253 (see Figure \ref{fig:mass_mass_diagram}). 
The maximum pulsar mass of $1.6 \, \rm M_{\odot}$ which corresponds to the minimum companion mass of $0.98 \, \rm M_{\odot}$ occurs for an inclination of 90$^\circ$, where the $\dot{\omega}$ constraint intersects with the mass function constraint. 
Using the measured value of $\dot{\omega}$ and assuming an isotropic distribution for the orbital inclination angle, we measured the probability distribution function of the pulsar and companion mass as shown in Figure~\ref{fig:mass_mass_diagram}. 

We find that for PSR J1325$-$6253, assuming that the pulsar has a mass between 1 and 2.5\,M$_\odot$, the pulsar mass is between 1.1 and 1.63\,$\rm M_{\odot}$ and the companion mass is between 0.98 and 1.47\,$\rm M_{\odot}$ (90\% confidence interval).
Figure \ref{fig:mass_mass_diagram} also shows the masses for DNS in which both NS masses are well measured. 
Their masses are tightly clustered, with recycled pulsar masses typically close to 1.3\,$\rm M_{\odot}$, extending up to 1.65\,$\rm M_{\odot}$. The masses of the non-recycled NSs are more evenly distributed between 1.15\,$\rm M_{\odot}$ and 1.5\,$\rm M_{\odot}$. Using these mass ranges as an astrophysical prior on the component masses for PSR J1325$-$6253, we find that the pulsar mass is 1.30--1.47\,$\rm M_{\odot}$, and the companion mass is 1.17--1.31\,$\rm M_{\odot}$, with an inclination of 49.0$^\circ$--57.4$^\circ$ (90 per cent confidence). \par

Given its low but significant eccentricity, its recycled nature and heavy companion mass, we have good reason to believe that the companion of PSR J1325$-$6253 is another (non-recycled) NS. If the companion of PSR J1325$-$6243 is a heavy white dwarf (WD) then there would be no supernova explosion hence; no mass-loss and kick. Therefore, like most other PSR--WD binaries, it would be expected to have an almost circular orbit ($e \lesssim 10^{-3}$) due to circularising effects of the preceding mass transfer phase that recycled the pulsar ~\citep{Phinney:1994ARA&A}. However, eccentric pulsar-white dwarf (PSR-WD) systems are no longer a theoretical conjecture. 
The discovery of PSR J1727$-$2946 \citep{lorimer_015} has filled the period and eccentricity gap between pulsars with NS and heavy WD companions. PSR J1727$-$2946 is a 27\,ms recycled pulsar with a heavy WD ($\sim 0.83 \, \rm M_{\odot}$) companion in a 40.3 d eccentric orbit ($e=0.0456$). Although this system may look similar to PSR J1325$-$6253, its large orbital period and low eccentricity (below the theoretical minimum eccentricity of $\sim 0.06$ for an ultra-stripped supernova, see section \ref{subsec:ultra_stripped?}) make it unrelated to any known DNS system (see Figure~\ref{fig:p_pb_corr} b). Apart from this, population synthesis models also disfavour the formation of DNS system with such orbital parameters \citep{Vigna-Gomez:2018dza}. Therefore, for PSR J1325$-$6253, a WD companion would require an evolutionary history in which the large orbital eccentricity ($e \sim 0.064$) has no natural explanation. For these reasons, we therefore conclude that the companion of PSR J1325$-$6253 is most likely an NS.

\section{Search for the companion} \label{sec:companion_search}

Given the likelihood that the companion of PSR J1325$-$6253 is an NS, the possibility exists that it might be detectable as a radio pulsar. 
In order to search for an NS companion of PSR J1325$-$6253, during each fold mode observation conducted using the UWL receiver and Medusa backend of the Parkes radio telescope, we also recorded the data in coherently dedispersed search mode and selected the 1000--1800\,MHz band\footnote{We selected this band because scattering at 1.6 GHz is relatively less than at 1.4 GHz and an additional 200 MHz bandwidth may increase the S/N of the pulsar (if present).} to perform the companion search. We dedispersed the time series at the given DM of the pulsar, and searched upto a maximum acceleration ($a_{\rm max}\simeq \pm 5 \rm m \, s^{-2}$) achieved by the system at its periastron using \texttt{PEASOUP}\footnote{\url{https://github.com/ewanbarr/peasoup}} which is a GPU accelerated search code and perform time-domain acceleration search. Candidates above spectral S/N of 5 were selected for folding and each folded candidate plot was inspected manually. No significant candidate was found using this method.\par

Since PSR J1325$-$6253 must be the first born NS as it has been recycled, the companion is expected to be a slow pulsar. 
Due to the presence of the red-noise at the higher end of the frequency power spectrum, FFT searches are less sensitive to slow pulsars, typically with  $P>1000$ ms. 
However, fast-folding algorithm (FFA) may overcome this difficulty and detect periodic signals which could have been masked during the dereddening process in FFT searches \citep[e.g.][]{cameron_017, parent_018, morello_020}. Therefore, in order to be sensitive to slow periods we used \texttt{RIPTIDE}\footnote{\url{https://github.com/v-morello/riptide}} which is an FFA based pulsar search code. To search with \texttt{RIPTIDE} we dedispersed time series at the known DM and created a dedispersed time series using \texttt{SIGPROC}\footnote{\url{https://github.com/SixByNine/sigproc}} and search for the periodic signals. However, in this case we did not resampled the time series as for slow pulsars e.g. J0737$-$3039B where the spin period of the pulsar is 2.77 sec. The loss in the sensitivity without acceleration search would be only $\sim 1 \%$ \citep{eatough13a} if the maximum acceleration of the system is $\pm 5 \rm \, ms^{-2}$. No convincing candidate was found using this method.\par

The accurately measured position of PSR J1325$-$6253 encouraged us to search for an optical counterpart in the third data release of SkyMapper's deep coverage of the Southern sky \citep{skymapper_019}. 
A main-sequence (MS) star comparable to the Sun, with a luminosity equal to the solar luminosity at a distance of 4.4--5.5\,kpc (as estimated through the DM for J1325$-$6253) would have an apparent magnitude of 18--18.5. The limiting magnitude of the Southern Sky Survey is $\sim$ 21 for all six filters ( $u,v,g,r,i$, and $z$) \citep{skymapper_07}, therefore an MS star should be easily visible. 
However, no visible counterpart was detected within 5--arcsec of the pulsar position in any of the six filters.
Therefore, we can rule out the possibility that the companion is a main-sequence star. 
Apart from this, a main sequence companion would also be inconsistent with the large orbital eccentricity and recycled nature of the pulsar, with the singular exception of PSR J1903$+$0327 which remains difficult to explain. 
This unusual system possesses a 2.15\,ms pulsar orbiting a solar-mass MS companion in a 95-d highly eccentric orbit of $e=0.44$ \citep{champion_08}.\par

The absence of any optical counterparts near PSR J1325$-6253$ also indicates that it is highly unlikely to detect any WD in this optical catalogue. Given the derived distance of 4--5\,kpc (corresponding to a DM of 303 $\rm pc \, cm^{-3}$) and the fact that massive WDs cool much faster than less massive ones, a putative white dwarf companion would likely be too faint to detect in optical observations even with no extinction \citep{agazie021}. We find that at such distances, the $V$-band magnitude of a WD would only be 26--27. Although this could potentially be achievable by extremely large optical telescopes with long exposure times, such a white dwarf is invisible in the existing catalogues.

\section{Discussion}
\label{sec:discussion}

It is interesting to explore whether this DNS system resembles others in the Milky Way. 
There are known correlations between the spin and orbital periods of the recycled pulsars in DNS systems~\citep{tauris17}. 
It has been argued, that in wider orbits, the accretion of matter onto the NS is shorter lived because the companion helium star is more evolved before it fills its Roche lobe \citep{Tauris:2015xra}. 
As a result, there should be a correlation between $P$ and $P_\mathrm{orb}$ of the observed DNS systems i.e., on average, pulsars in wider orbits should have larger $P$. 
Such a correlation was first observed by \citet{tauris17} and a fit is given by,

\begin{equation}
\label{eq:p_pb_relation}
P = 44 \, \textrm{ms} \, (P_{\rm{orb}}/\rm days)^{0.26} .
\end{equation}

\begin{figure}
\centering
\includegraphics[width=0.45\textwidth]{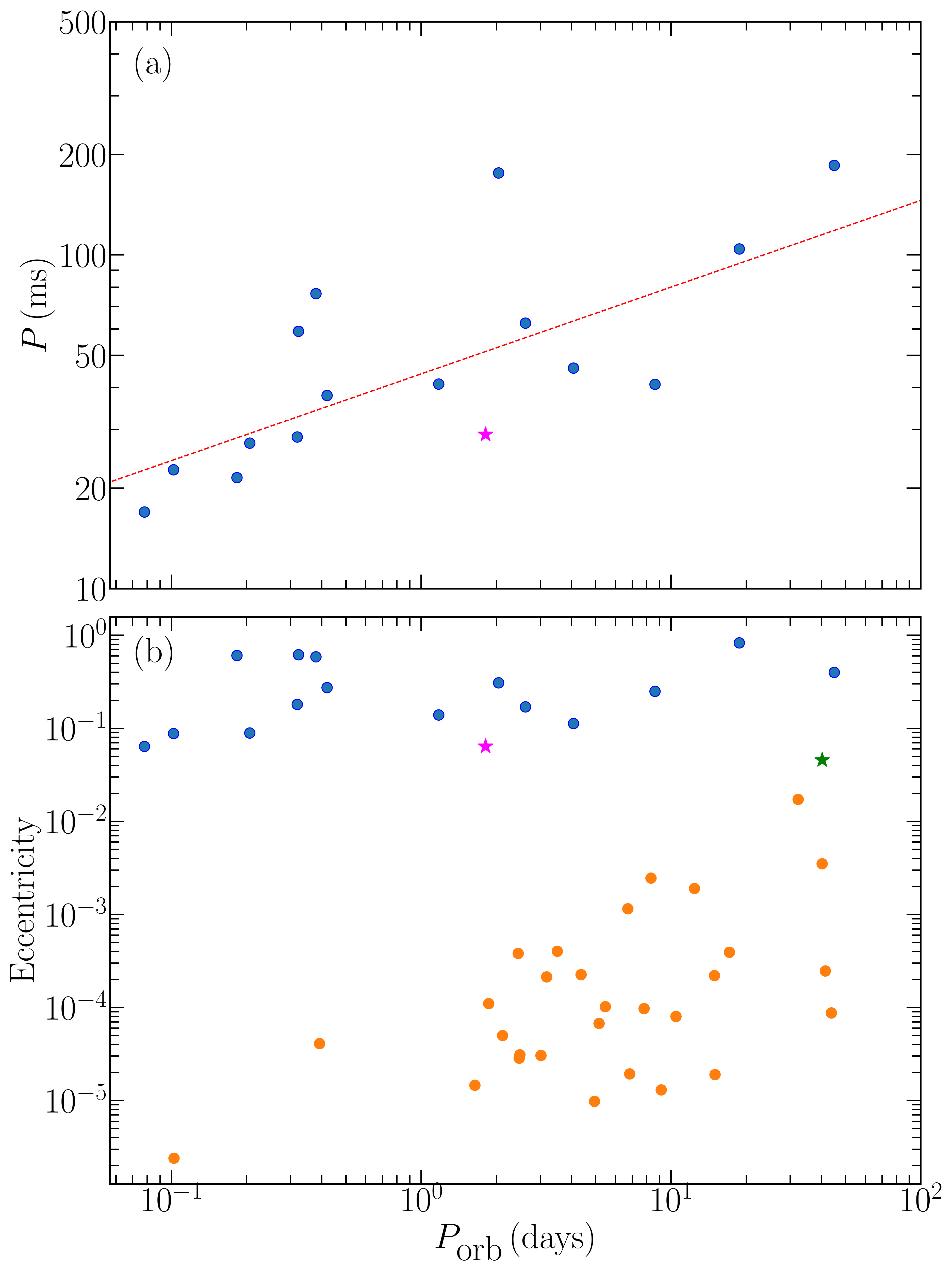}
\caption{Correlations between the orbital period $P_{\rm orb}$ of recycled DNS systems (blue) and the spin period of the pulsars (top panel) and the orbital eccentricity (bottom panel). PSR J1325$-$6253 is shown as a pink star. The red dashed line in the $P_{\rm orb}-P$ plot (top panel) is the relation from \citet{tauris17}. We also show the orbital periods and eccentricities for PSR-WD binaries (orange) with $P > 15$\,ms and $P_{\textrm{orb}} < 100$\,d in the bottom panel. The green star is PSR J1727$-$2946 \citep{lorimer_015} which is discussed in the text in Section~\ref{sec:system_nature}.}
\label{fig:p_pb_corr}
\end{figure}

This $P_{\rm orb}$--$P$ correlation is shown in Figure~\ref{fig:p_pb_corr}(a) where short orbital period DNS systems tend to have shorter spin periods, and PSR J1325$-$6253 follows this trend, reinforcing the notion that it shared a common evolutionary history with the other DNSs.

Following similar arguments, for pulsars in wider orbits, less material is transferred to the first born NS with increasing $P_{\rm orb}$ and on explosion a higher mass is ejected which increases the eccentricity of the system. 
Therefore, a weak correlation between $P_{\rm orb}$ and $e$ should also exist \citep{tauris17}. 
In Figure~\ref{fig:p_pb_corr}(b), we have shown the updated $P_{\rm orb}$-$e$ distribution of the known recycled DNS systems in the Galactic disc. 
The evidence for this correlation is now confirmed to be weak as the correlation coefficient for the $P_{\rm orb}$-$e$ distribution is just 0.28 with $p$-value $>0.05$, but there does appear to be a bifurcation of the population into low-eccentricity ($e<0.3$) and high-eccentricity ($e>0.5$) systems \citep[][]{Andrews:2019vou}, for each of which one could argue that the relation still holds. For example, the low-eccentricity systems alone show a correlation coefficient of 0.81.

\begin{figure}
\centering
\includegraphics[width=0.45\textwidth]{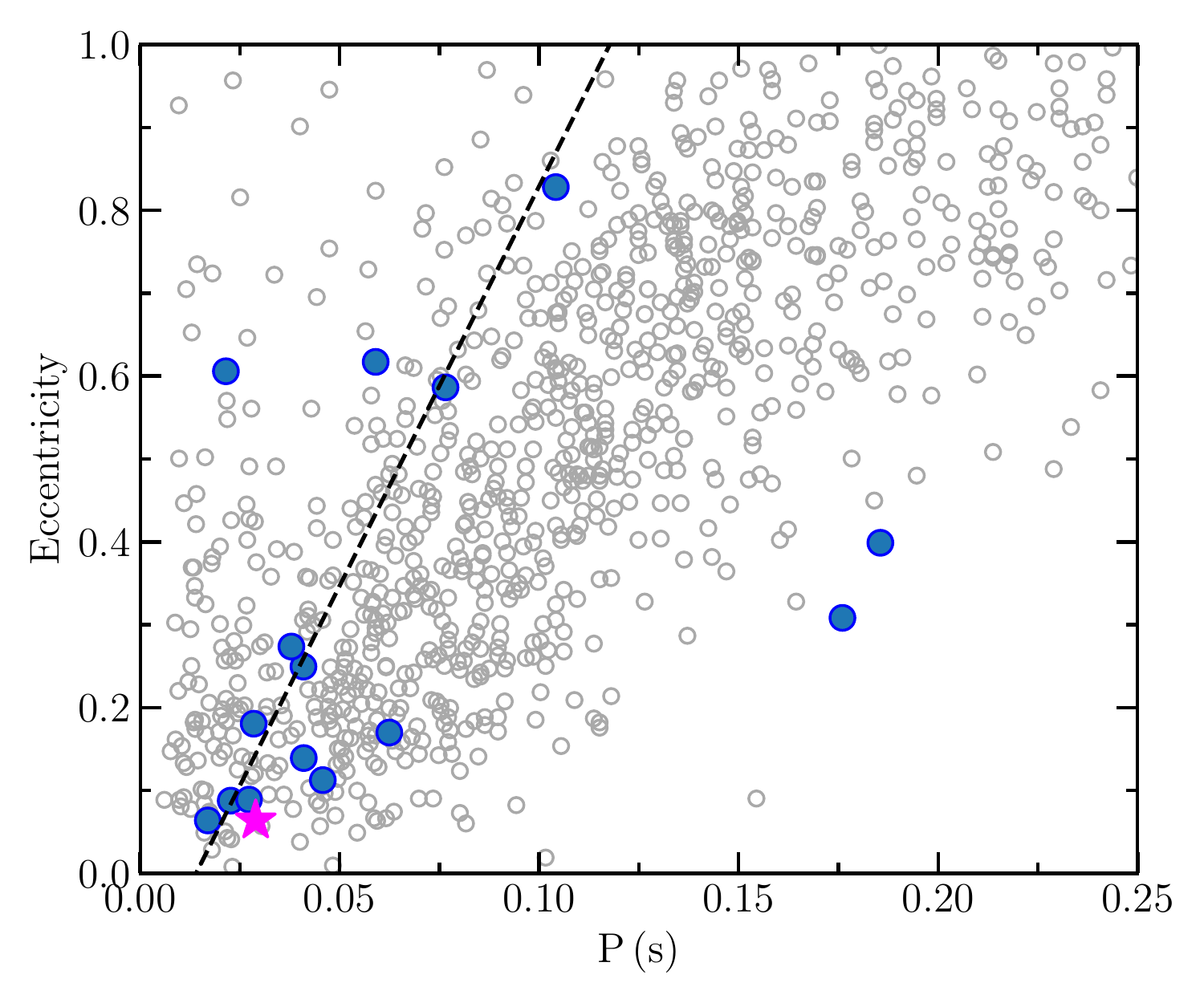}
\caption{The $P$--$e$ distribution of known recycled DNS systems (in blue) and simulated DNS systems (in grey) from \citet{dewi05}. The data points of the simulations were obtained by accurately digitising Figure 3(d) in \citet{dewi05}. Obviously the model does not accurately predict the parameters of the observed population. The dashed line represents an updated linear fit to 11 DNS systems. This relation was first obtained by \citet{faulkner04} using 7 DNS systems.}
\label{fig:p_ecc}
\end{figure}

There is also an observed relation between the spin period of the recycled pulsar and the orbital eccentricity of the non-globular cluster DNSs as first seen by \cite{faulkner04}. A strong correlation between $P$ and $e$ was first observed by \cite{faulkner04} with a Pearson coefficient of 0.97. PSR J1325$-$6253 also follows this trend as shown in Figure~\ref{fig:p_ecc}. Later, \cite{dewi05} simulated DNS systems in the $P$--$e$ phase space and found that the $P$--$e$ relation is consistent for DNS systems if they experienced low kicks ($\sigma<50 \, \textrm{km  s}^{-1}$) during the second supernova explosion. In Figure~\ref{fig:p_ecc}, we show the updated best-fitting relation to the known DNS systems and simulated DNS systems from \cite{dewi05} in grey circles in $P$--$e$ space. Clearly the observed DNS distribution is not well fit by the simulations, although the tendency for longer orbital period systems to have longer spin periods remains valid.

\begin{figure}
\centering
\includegraphics[width=0.45\textwidth]{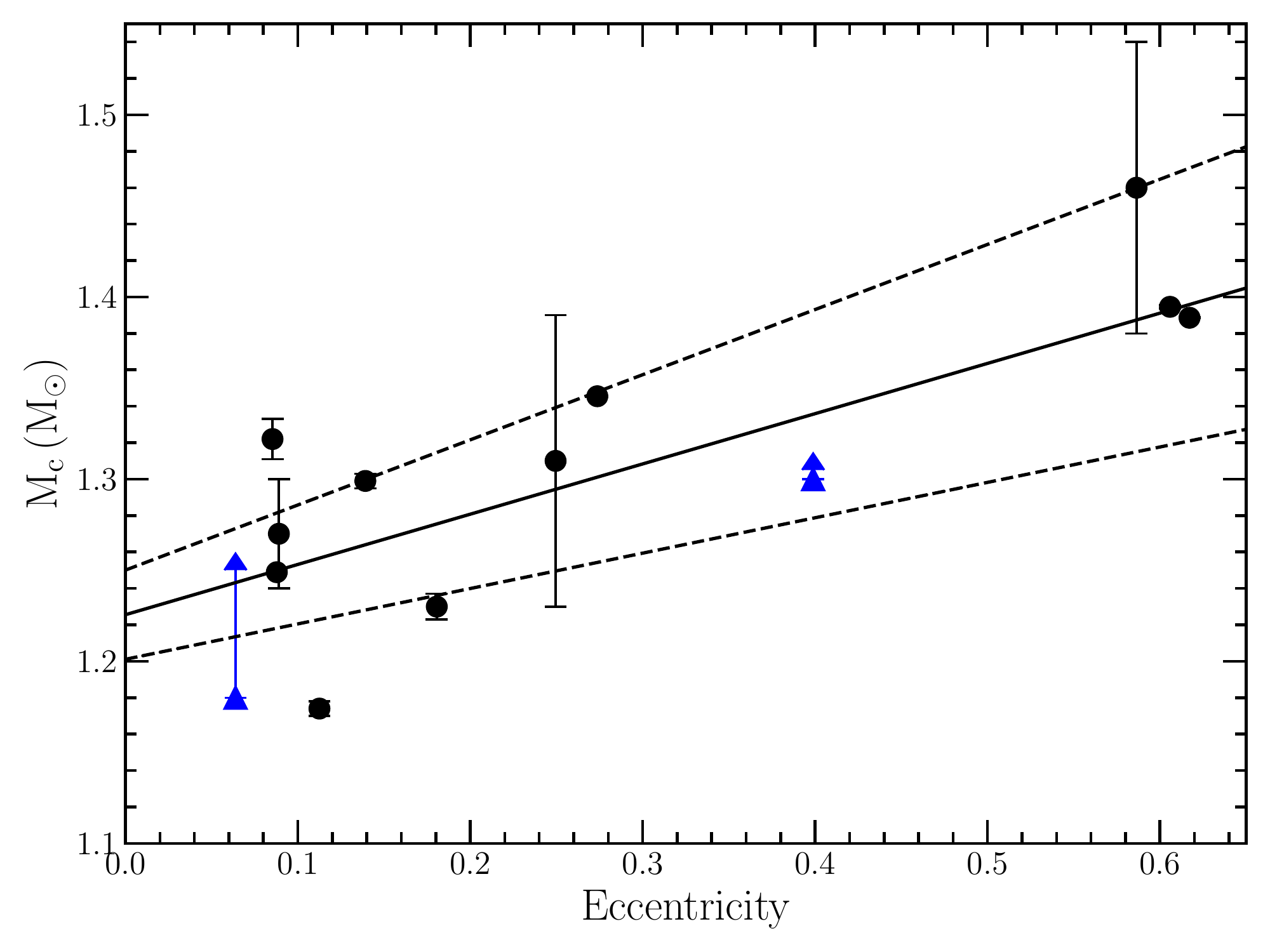}
\caption{Companion mass vs eccentricity for DNS systems. This plot includes DNS systems for which the companion mass is well measured (in black) along with their error bars and three systems (in blue) for which the minimum  companion mass ($> 1.17 M_{\odot}$) has been reported assuming an orbital inclination angle, $i=90^{\circ}$. For these systems the upper limit on companion mass is the average value of the total system mass. Solid black line correspond to the best fit line using MCMC and dashed black lines are lines within 1-sigma error bars.}
\label{fig:mc_ecc}
\end{figure}

\subsection{A case of an ultra-stripped supernova?}
\label{subsec:ultra_stripped?}

It has been suggested that DNS systems with low eccentricities are the result of ultra-stripped electron-capture SNe in which, instead of the usual iron-core collapse, the collapse is triggered when O--Ne--Mg core starts capturing electrons onto Mg and Ne \citep{nomoto84,podsiadlowski04}. 
The mass of the core before the collapse is just 1.37--1.47 $\rm M_{\odot}$ and ejecta mass is $<0.2 \, \textrm{M}_{\odot}$, so that low-amplitude kicks are imparted during the second SN \citep{suwa015, muller16, tauris17}. This new formation channel of second born NSs in DNS system is able to explain the low eccentricities in some of the DNS systems. \cite{Tauris:2015xra} found that small kicks occur for the lowest mass NSs, and that theoretically a correlation between NS mass and kicks at birth should exist, which then implies a correlation between NS mass and the observed eccentricity ($M_{c}$--$e$). \cite{tauris17} first observed a hint of a $M_{c}$-$e$ correlation using six DNS systems with well-constrained companion masses. 
In Figure \ref{fig:mc_ecc}, we have plotted the updated $M_{c}$--$e$ distribution of 10 DNS systems for which the companion mass is well constrained. 
We have also included two other DNS systems for which the minimum companion mass assuming orbital inclination, $i=90^{\circ}$ is above $1.17 \, \textrm{M}_{\odot}$. Given the large error bars on the companion masses for some of these systems, we used a Monte-Carlo method to find the best fitting relation. We find a significant correlation with a Pearson coefficient of 0.82 and a significance greater than 99.99 per cent. 
The observed scatter about this best fitting relation could be due to different kick directions at birth. Among the known DNS system population PSRs J0737$-$3039, J0453$+$1559 and J1411$+$2551 are considered to have originated from ultra-stripped SNe. Based on simulations, the upper limits on the kick velocities for both J0453$+$1559 and J1411$+$2551 are  $<100$\,km\,s$^{-1}$, consistent with low kicks during the second supernova~\citep{tauris17}. \par

In the case of PSR J1325$-$6253, the low eccentricity of only 0.064, the wide orbit and the low estimated companion mass are in agreement with the hypothesis that it was formed via an ultra-stripped SN. In the absence of any dynamical effects, the pre-ultra-stripped SN orbital periods of the binary are typically lie in range of 1\,hr-2\,d \citep{Tauris:2015xra}. The existing orbital period of 1.81\,d and low eccentricity for J1325$-$6253 mean that prior to the explosion the orbital period was probably only marginally shorter. Using the relation of binary coalescence time, $\tau_\mathrm{GW}$ due to gravitational damping from \cite{peters64}, we found it to be 212\,Gyr, implying that the eccentricity and orbital period of PSR J1325$-$6253 have not been affected by gravitational damping to any significance, and thus that the system currently resembles its post ultra-stripped SN state. \par

As discussed above, the kicks imparted to NSs during ultra-stripped SN are expected to be weaker compared to more typical iron-core collapse SNe. 
For a symmetric ultra-stripped SN, no kick is imparted on the second born SN. 
The post-SN eccentricity in such cases is then proportional to the mass-loss i.e., $e=\Delta M/M_{\textrm{sys}}$.
There are two main contributions to the mass-loss $\Delta M$. 
The first contribution is from any mass ejected from the outer layers of the star (e.g., the helium envelope, $M_{\textrm{He,f}}^{\textrm{env}}$) during the supernova.
Additionally, around 99\% of the energy of the supernova is emitted in the form of neutrinos during the formation of the NS.
The gravitational binding energy of a NS depends in detail on the NS equation of state, but is approximated for a range of EOSs by $\Delta M_{\textrm{NS}}^{\textrm{bind}} = 0.084\, \textrm{M}_{\odot} \cdot (M_{\textrm{NS}}/\textrm{M}_{\odot})^{2}$ \citep{lattimer_and_yahil89,Lattimer:2021review}. 
\cite{Tauris:2015xra} have an electron-capture SN model which is similar to the orbital period of PSR J1325$-$6253, in which the initial orbital period of the binary is 2.6 days and the final orbital period prior to the supernova explosion is 1.78-d. 
In this model, the final core mass is $M_{\textrm{core, f}}=1.37 \, \textrm{M}_{\odot}$ and $M_{\textrm{He,f}}^{\textrm{env}} \simeq 0.08 \, \textrm{M}_{\odot}$ i.e., the total mass of the helium star, $M_{\rm f}$ is $1.45 \textrm{M}_{\odot}$ just before the explosion. \par

As discussed above, the released binding energy depends on the $M_{\textrm{NS}}^{2}$ which is not well constrained in our case unless we invoke an astrophysical prior based upon known recycled pulsar masses. Using the probability distribution for the companion mass from Section~\ref{sec:properties}, we find that the released binding energy is between 0.08 and 0.18\,M$_\odot$. If we assume the astrophysical prior on the NS mass (as shown in Figure~\ref{fig:mass_mass_diagram}), the companion mass is better constrained, and the mass-loss is constrained between 0.11 and 0.14\,M$_\odot$ (all values 90\% confidence). This mass-loss implies an orbital eccentricity of 0.03--0.07 (0.04--0.06) under the two priors. The observed eccentricity ($e = 0.064$) falls at the 91$^\mathrm{st}$ percentile of the distribution for the isotropic inclination prior, and has no support under the astrophysical prior. This implies that mass-loss due to neutrino emission is insufficient to explain the observed eccentricity. The observed eccentricity corresponds to a mass-loss of $\Delta M = 0.16$--$0.17$\,M$_\odot$. Thus, an additional $0.02$--$0.09$\,M$_\odot$ (0.03--0.06\,M$_\odot$) of mass-loss is required to explain the observed eccentricity. This additional mass-loss is likely to be associated with mass ejected from the star during the supernova. This mass-loss is less than the $0.08$\,M$_\odot$ of helium envelope found in the model of \citet{Tauris:2015xra} discussed above.

While the low eccentricity may be suggestive of an ultra-stripped supernova with little or no ejecta, some NSs are known to have kicks of hundreds $\rm km\,s^{-1}$ \citep[e.g.,][]{Hobbs:2005yx}, and substantial mass-loss. The impact of relatively large kicks on the binary orbit can be compensated by increased mass-loss, and vice versa. Therefore the most general solution allows both of these to be free parameters. 
In order to constrain the kick and mass-loss for PSR J1325$-$6253, we simulated $\sim 10^{7}$ secondary supernova explosions, broadly following the method of \citet{tauris17}. We assume that the orbit is initially circular prior to the supernova, as is expected from binary evolution models.
We assume that NS kicks are isotropically distributed in the rest frame of the exploding star, and assume a uniform distribution for both the magnitude of the kick and the amount of ejected mass. We then constrain these parameters by selecting only those systems that reproduce the orbital period and eccentricity of PSR J1325$-$6253 following the supernova. The magnitude of the kick is constrained to be $<180$\,km\,s$^{-1}$, and the ejecta mass is $<1.6$\, M$_{\odot}$ with 90$\%$ confidence. The distributions of kick magnitude and ejecta mass peak at $20$\,km\,s$^{-1}$ and 0.15\,$\rm M_{\odot}$ respectively. Whilst still fairly broad, these constraints are consistent with what we would expect from an ultra-stripped supernova.

\section{Conclusion}
\label{sec:conclusion}

We have presented the discovery and timing of the binary pulsar, PSR J1325$-$6253, from a reprocessing of the HTRU-S LowLat survey. Its location in $P$--$\dot{P}$ space suggests that the pulsar is recycled. PSR J1325$-$6253 is in a binary with an orbital period of 1.81\,d. Based on the evidence for recycling, the orbital eccentricity ($e = 0.064$) and the high total mass ($M_{\rm tot} = 2.57 \pm 0.06 $\,M$_\odot$) we argue that the companion of PSR J1325$-$6253 is likely another NS, with a low mass of $\sim 1.2$\,M$_\odot$ (Figure~\ref{fig:mass_mass_diagram}). Dedicated timing of this system using the Parkes radio-telescope allowed us to measure one of the post-Keplerian parameters ($\dot{\omega}$) with high significance. Its wide orbit (1.81\,d) excludes the possibility of measuring other post-Keplerian parameters in the near future, which is required to uniquely determine the system 
component masses. The Shapiro delay has not been detected in this system, which suggests that the orbital inclination is unlikely to be high and nearly edge-on. However, given the current 42\,$\mu$s timing residuals, we would not expect to detect the Shapiro delay using Parkes radio telescope, as the range $r$ of the Shapiro delay would be $\sim 6.15$\,$\mu$s if we assume the companion mass is $1.25$\, M$_{\odot}$. However, the 64-dish MeerKat radio telescope is a new highly sensitive radio telescope in South Africa that provides a factor of about six improvement in sensitivity over the Parkes telescope at 20cm and is well suited for measuring additional relativistic parameters in this pulsar \citep{bailes_20,kramer_21b}. A dedicated campaign that covered the entire orbital phase could help determine the amplitude of the Shapiro delay and hence the masses of the pulsar and companion. With continued precision timing, in the future, we hope to measure its proper motion which would be helpful to place tighter constraints on the mass-loss and kick from the second supernova, although at the likely distance to the source, a parallax measurement would be extremely challenging. Without an accurate distance, the proper motion will not be overly constraining. The most compelling characteristic of PSR J1325$-$6253 is its low eccentricity which is the lowest among any known DNS system in a wide orbit, and second lowest after PSR J1946+2052 \citep{stovall018} although in this system there is likely to have been significant decay of the eccentricity due to relativistic decay of the orbit since the final explosion. 
We argue that our system represents one of the most extreme cases of an ultra-stripped supernova \citep{Tauris:2015xra}.

The discovery of the Crab pulsar inside its young supernova remnant biased our view of ``typical'' supernova explosions in which many solar masses are ejected, and the remnant is lit up by the powerful pulsar and visible for thousands of years.  
PSR J1325$-$6253's final supernova explosion would have been in stark comparison to the Crab, with little baryonic mass ejection. 
It would have a very low optical luminosity both after its initial explosion and a very small lifetime as a visible supernova remnant.
Such an example may be iPTF 14gqr (SN 2014ft), recently detected by the intermediate Palomar Transient Factory \citep{de_018}. 
The SN faded from view in just a few days and had an optical magnitude of $\sim$20.2 mag in $g$-band at redshift $z$ = 0.063.
It is also interesting to postulate whether the tiny ejecta  associated with such ultra-stripped SNe may permit the creation and detection of prompt radio emission, normally shielded by the many solar masses of ejecta. Such emission may mirror that of one-off fast radio bursts, rare millisecond-duration pulses visible at cosmological distances \citep{lorimer2007,thorton_013}, although the galactic birthrate of  ultra-stripped SNe ($\leq10^{-4}$ yr$^{-1}$ ) is insufficient to account for all of them.

\section*{Acknowledgments}

We thank Thomas M. Tauris and anonymous referee for useful comments to improve the article. We also thank Paul Lasky for providing insights about the NS masses and EOSs. RS and SS are supported by the Australian Research Council (ARC) Centre of Excellence for Gravitational Wave Discovery OzGrav, through project number CE170100004. SS is a recipient of an ARC Discovery Early Career Research Award (DE220100241).
The reprocessing of the HTRU-S LowLat survey made an extensive use of the GPU accelerators on OzStar supercomputer at Swinburne University of Technology. The OzSTAR program receives funding in part from the Astronomy National Collaborative Research Infrastructure Strategy (NCRIS) allocation provided by the Australian Government. This work would not have been possible without the 64-m Parkes radio telescope, the dedication and professionalism of the staff, and the capabilities of the UWL receiver and MEDUSA backend. The Parkes radio telescope is part of the Australia Telescope National Facility (grid.421683.a) which is funded by the Australian Government for operation as a National Facility managed by CSIRO. We acknowledge the Wiradjuri people as the traditional owners of the Observatory site. The localisation of the source was completed using the TRAPUM backend on the 64-dish MeerKAT radio telescope owned and operated by SARAO.

\section*{Data Availability}

The data used for this article will be shared on reasonable request to the corresponding author.


\bibliographystyle{mnras}
\bibliography{dns_mnras} 

\begin{thebibliography}{}
\makeatletter
\relax
\def\mn@urlcharsother{\let\do\@makeother \do\$\do\&\do\#\do\^\do\_\do\%\do\~}
\def\mn@doi{\begingroup\mn@urlcharsother \@ifnextchar [ {\mn@doi@}
  {\mn@doi@[]}}
\def\mn@doi@[#1]#2{\def\@tempa{#1}\ifx\@tempa\@empty \href
  {http://dx.doi.org/#2} {doi:#2}\else \href {http://dx.doi.org/#2} {#1}\fi
  \endgroup}
\def\mn@eprint#1#2{\mn@eprint@#1:#2::\@nil}
\def\mn@eprint@arXiv#1{\href {http://arxiv.org/abs/#1} {{\tt arXiv:#1}}}
\def\mn@eprint@dblp#1{\href {http://dblp.uni-trier.de/rec/bibtex/#1.xml}
  {dblp:#1}}
\def\mn@eprint@#1:#2:#3:#4\@nil{\def\@tempa {#1}\def\@tempb {#2}\def\@tempc
  {#3}\ifx \@tempc \@empty \let \@tempc \@tempb \let \@tempb \@tempa \fi \ifx
  \@tempb \@empty \def\@tempb {arXiv}\fi \@ifundefined
  {mn@eprint@\@tempb}{\@tempb:\@tempc}{\expandafter \expandafter \csname
  mn@eprint@\@tempb\endcsname \expandafter{\@tempc}}}

\bibitem[\protect\citeauthoryear{{Abbott} et~al.,}{{Abbott}
  et~al.}{2017a}]{abbott17}
{Abbott} B.~P.,  et~al., 2017a, \mn@doi [\prl]
  {10.1103/PhysRevLett.119.161101}, \href
  {https://ui.adsabs.harvard.edu/abs/2017PhRvL.119p1101A} {119, 161101}

\bibitem[\protect\citeauthoryear{{Abbott} et~al.,}{{Abbott}
  et~al.}{2017b}]{LIGOScientific:2017zic}
{Abbott} B.~P.,  et~al., 2017b, \mn@doi [\apjl] {10.3847/2041-8213/aa920c},
  \href {https://ui.adsabs.harvard.edu/abs/2017ApJ...848L..13A} {848, L13}

\bibitem[\protect\citeauthoryear{Abbott et~al.,}{Abbott
  et~al.}{2020}]{LIGOScientific:2020aai}
Abbott B.,  et~al., 2020, \mn@doi [\apjl] {10.3847/2041-8213/ab75f5}, \href
  {https://ui.adsabs.harvard.edu/abs/2020ApJ...892L...3A} {892, L3}

\bibitem[\protect\citeauthoryear{{Agazie} et~al.,}{{Agazie}
  et~al.}{2021}]{agazie021}
{Agazie} G.~Y.,  et~al., 2021, \mn@doi [\apj] {10.3847/1538-4357/ac142b}, \href
  {https://ui.adsabs.harvard.edu/abs/2021ApJ...922...35A} {922, 35}

\bibitem[\protect\citeauthoryear{{Andrews} \& {Mandel}}{{Andrews} \&
  {Mandel}}{2019}]{Andrews:2019vou}
{Andrews} J.~J.,  {Mandel} I.,  2019, \mn@doi [\apjl]
  {10.3847/2041-8213/ab2ed1}, \href
  {https://ui.adsabs.harvard.edu/abs/2019ApJ...880L...8A} {880, L8}

\bibitem[\protect\citeauthoryear{{Bailes} et~al.,}{{Bailes}
  et~al.}{2020}]{bailes_20}
{Bailes} M.,  et~al., 2020, \mn@doi [\pasa] {10.1017/pasa.2020.19}, \href
  {https://ui.adsabs.harvard.edu/abs/2020PASA...37...28B} {37, e028}

\bibitem[\protect\citeauthoryear{{Belczynski}, {Kalogera}, {Rasio}, {Taam},
  {Zezas}, {Bulik}, {Maccarone}  \& {Ivanova}}{{Belczynski}
  et~al.}{2008}]{belczynski_08}
{Belczynski} K.,  {Kalogera} V.,  {Rasio} F.~A.,  {Taam} R.~E.,  {Zezas} A.,
  {Bulik} T.,  {Maccarone} T.~J.,   {Ivanova} N.,  2008, \mn@doi [\apjs]
  {10.1086/521026}, \href
  {https://ui.adsabs.harvard.edu/abs/2008ApJS..174..223B} {174, 223}

\bibitem[\protect\citeauthoryear{Berti et~al.,}{Berti et~al.}{2015}]{berti_15}
Berti E.,  et~al., 2015, \mn@doi [Classical and Quantum Gravity]
  {10.1088/0264-9381/32/24/243001}, 32, 243001

\bibitem[\protect\citeauthoryear{{Burgay} et~al.,}{{Burgay}
  et~al.}{2003}]{burgay_03}
{Burgay} M.,  et~al., 2003, \mn@doi [\nat] {10.1038/nature02124}, \href
  {https://ui.adsabs.harvard.edu/abs/2003Natur.426..531B} {426, 531}

\bibitem[\protect\citeauthoryear{{Burgay} et~al.,}{{Burgay}
  et~al.}{2006}]{phps_06}
{Burgay} M.,  et~al., 2006, \mn@doi [\mnras]
  {10.1111/j.1365-2966.2006.10100.x}, \href
  {https://ui.adsabs.harvard.edu/abs/2006MNRAS.368..283B} {368, 283}

\bibitem[\protect\citeauthoryear{{Cameron}, {Barr}, {Champion}, {Kramer}  \&
  {Zhu}}{{Cameron} et~al.}{2017}]{cameron_017}
{Cameron} A.~D.,  {Barr} E.~D.,  {Champion} D.~J.,  {Kramer} M.,   {Zhu} W.~W.,
   2017, \mn@doi [\mnras] {10.1093/mnras/stx589}, \href
  {https://ui.adsabs.harvard.edu/abs/2017MNRAS.468.1994C} {468, 1994}

\bibitem[\protect\citeauthoryear{{Cameron} et~al.,}{{Cameron}
  et~al.}{2018}]{cameronJ1757018}
{Cameron} A.~D.,  et~al., 2018, \mn@doi [\mnras] {10.1093/mnrasl/sly003}, \href
  {https://ui.adsabs.harvard.edu/abs/2018MNRAS.475L..57C} {475, L57}

\bibitem[\protect\citeauthoryear{{Cameron} et~al.,}{{Cameron}
  et~al.}{2020}]{cameron_020}
{Cameron} A.~D.,  et~al., 2020, \mn@doi [\mnras] {10.1093/mnras/staa039}, \href
  {https://ui.adsabs.harvard.edu/abs/2020MNRAS.493.1063C} {493, 1063}

\bibitem[\protect\citeauthoryear{{Champion} et~al.,}{{Champion}
  et~al.}{2005}]{champion_05}
{Champion} D.~J.,  et~al., 2005, \mn@doi [\mnras]
  {10.1111/j.1365-2966.2005.09499.x}, \href
  {https://ui.adsabs.harvard.edu/abs/2005MNRAS.363..929C} {363, 929}

\bibitem[\protect\citeauthoryear{{Champion} et~al.,}{{Champion}
  et~al.}{2008}]{champion_08}
{Champion} D.~J.,  et~al., 2008, \mn@doi [Science] {10.1126/science.1157580},
  \href {https://ui.adsabs.harvard.edu/abs/2008Sci...320.1309C} {320, 1309}

\bibitem[\protect\citeauthoryear{Chattopadhyay, Stevenson, Hurley, Rossi  \&
  Flynn}{Chattopadhyay et~al.}{2020}]{Chattopadhyay:2019xye}
Chattopadhyay D.,  Stevenson S.,  Hurley J.~R.,  Rossi L.~J.,   Flynn C.,
  2020, \mn@doi [Mon. Not. Roy. Astron. Soc.] {10.1093/mnras/staa756}, 494,
  1587

\bibitem[\protect\citeauthoryear{{Chaurasia} \& {Bailes}}{{Chaurasia} \&
  {Bailes}}{2005}]{chaurasia05}
{Chaurasia} H.~K.,  {Bailes} M.,  2005, \mn@doi [\apj] {10.1086/444447}, \href
  {https://ui.adsabs.harvard.edu/abs/2005ApJ...632.1054C} {632, 1054}

\bibitem[\protect\citeauthoryear{Cordes \& Lazio}{Cordes \&
  Lazio}{2002}]{ne2001}
Cordes J.~M.,  Lazio T. J.~W.,  2002, arXiv preprint astro-ph/0207156

\bibitem[\protect\citeauthoryear{{Cordes} et~al.,}{{Cordes}
  et~al.}{2006}]{alfa_06}
{Cordes} J.~M.,  et~al., 2006, \mn@doi [\apj] {10.1086/498335}, \href
  {https://ui.adsabs.harvard.edu/abs/2006ApJ...637..446C} {637, 446}

\bibitem[\protect\citeauthoryear{{Corongiu}, {Kramer}, {Stappers}, {Lyne},
  {Jessner}, {Possenti}, {D'Amico}  \& {L{\"o}hmer}}{{Corongiu}
  et~al.}{2007}]{corongiu07}
{Corongiu} A.,  {Kramer} M.,  {Stappers} B.~W.,  {Lyne} A.~G.,  {Jessner} A.,
  {Possenti} A.,  {D'Amico} N.,   {L{\"o}hmer} O.,  2007, \mn@doi [\aap]
  {10.1051/0004-6361:20054385}, \href
  {https://ui.adsabs.harvard.edu/abs/2007A&A...462..703C} {462, 703}

\bibitem[\protect\citeauthoryear{{Damour} \& {Deruelle}}{{Damour} \&
  {Deruelle}}{1986}]{damour_86}
{Damour} T.,  {Deruelle} N.,  1986, Ann. Inst. Henri Poincar{\'e} Phys.
  Th{\'e}or, \href {https://ui.adsabs.harvard.edu/abs/1986AIHS...44..263D} {44,
  263}

\bibitem[\protect\citeauthoryear{{De} et~al.,}{{De} et~al.}{2018}]{de_018}
{De} K.,  et~al., 2018, \mn@doi [Science] {10.1126/science.aas8693}, \href
  {https://ui.adsabs.harvard.edu/abs/2018Sci...362..201D} {362, 201}

\bibitem[\protect\citeauthoryear{{Dewi}, {Podsiadlowski}  \& {Pols}}{{Dewi}
  et~al.}{2005}]{dewi05}
{Dewi} J.~D.~M.,  {Podsiadlowski} P.,   {Pols} O.~R.,  2005, \mn@doi [\mnras]
  {10.1111/j.1745-3933.2005.00085.x}, \href
  {https://ui.adsabs.harvard.edu/abs/2005MNRAS.363L..71D} {363, L71}

\bibitem[\protect\citeauthoryear{{Eatough}, {Kramer}, {Lyne}  \&
  {Keith}}{{Eatough} et~al.}{2013}]{eatough13a}
{Eatough} R.~P.,  {Kramer} M.,  {Lyne} A.~G.,   {Keith} M.~J.,  2013, \mn@doi
  [\mnras] {10.1093/mnras/stt161}, \href
  {https://ui.adsabs.harvard.edu/abs/2013MNRAS.431..292E} {431, 292}

\bibitem[\protect\citeauthoryear{Farrow, Zhu  \& Thrane}{Farrow
  et~al.}{2019}]{Farrow:2019xnc}
Farrow N.,  Zhu X.-J.,   Thrane E.,  2019, \mn@doi [Astrophys. J.]
  {10.3847/1538-4357/ab12e3}, 876, 18

\bibitem[\protect\citeauthoryear{{Faulkner} et~al.,}{{Faulkner}
  et~al.}{2004}]{faulkner04}
{Faulkner} A.~J.,  et~al., 2004, \mn@doi [\mnras]
  {10.1111/j.1365-2966.2004.08310.x}, \href
  {https://ui.adsabs.harvard.edu/abs/2004MNRAS.355..147F} {355, 147}

\bibitem[\protect\citeauthoryear{{Ferdman} \& {PALFA Collaboration}}{{Ferdman}
  \& {PALFA Collaboration}}{2018}]{ferdman_018}
{Ferdman} R.~D.,  {PALFA Collaboration} 2018, in {Weltevrede} P.,  {Perera}
  B.~B.~P.,  {Preston} L.~L.,   {Sanidas} S.,  eds,  Vol. 337, Pulsar
  Astrophysics the Next Fifty Years. pp 146--149,
  \mn@doi{10.1017/S1743921317009139}

\bibitem[\protect\citeauthoryear{{Ferdman} et~al.,}{{Ferdman}
  et~al.}{2014}]{ferdman014}
{Ferdman} R.~D.,  et~al., 2014, \mn@doi [\mnras] {10.1093/mnras/stu1223}, \href
  {https://ui.adsabs.harvard.edu/abs/2014MNRAS.443.2183F} {443, 2183}

\bibitem[\protect\citeauthoryear{{Ferdman} et~al.,}{{Ferdman}
  et~al.}{2020}]{ferdman2020}
{Ferdman} R.~D.,  et~al., 2020, \mn@doi [\nat] {10.1038/s41586-020-2439-x},
  \href {https://ui.adsabs.harvard.edu/abs/2020Natur.583..211F} {583, 211}

\bibitem[\protect\citeauthoryear{{Fonseca}, {Stairs}  \& {Thorsett}}{{Fonseca}
  et~al.}{2014}]{fonseca014}
{Fonseca} E.,  {Stairs} I.~H.,   {Thorsett} S.~E.,  2014, \mn@doi [\apj]
  {10.1088/0004-637X/787/1/82}, \href
  {https://ui.adsabs.harvard.edu/abs/2014ApJ...787...82F} {787, 82}

\bibitem[\protect\citeauthoryear{Galaudage, Adamcewicz, Zhu, Stevenson  \&
  Thrane}{Galaudage et~al.}{2021}]{Galaudage:2020zst}
Galaudage S.,  Adamcewicz C.,  Zhu X.-J.,  Stevenson S.,   Thrane E.,  2021,
  \mn@doi [Astrophys. J. Lett.] {10.3847/2041-8213/abe7f6}, 909, L19

\bibitem[\protect\citeauthoryear{{Haniewicz}, {Ferdman}, {Freire}, {Champion},
  {Bunting}, {Lorimer}  \& {McLaughlin}}{{Haniewicz}
  et~al.}{2021}]{Haniewicz_21}
{Haniewicz} H.~T.,  {Ferdman} R.~D.,  {Freire} P.~C.~C.,  {Champion} D.~J.,
  {Bunting} K.~A.,  {Lorimer} D.~R.,   {McLaughlin} M.~A.,  2021, \mn@doi
  [\mnras] {10.1093/mnras/staa3466}, \href
  {https://ui.adsabs.harvard.edu/abs/2021MNRAS.500.4620H} {500, 4620}

\bibitem[\protect\citeauthoryear{{Hobbs}, {Lorimer}, {Lyne}  \&
  {Kramer}}{{Hobbs} et~al.}{2005}]{Hobbs:2005yx}
{Hobbs} G.,  {Lorimer} D.~R.,  {Lyne} A.~G.,   {Kramer} M.,  2005, \mn@doi
  [\mnras] {10.1111/j.1365-2966.2005.09087.x}, \href
  {https://ui.adsabs.harvard.edu/abs/2005MNRAS.360..974H} {360, 974}

\bibitem[\protect\citeauthoryear{{Hobbs} et~al.,}{{Hobbs} et~al.}{2020}]{uwl}
{Hobbs} G.,  et~al., 2020, \mn@doi [\pasa] {10.1017/pasa.2020.2}, \href
  {https://ui.adsabs.harvard.edu/abs/2020PASA...37...12H} {37, e012}

\bibitem[\protect\citeauthoryear{{Hotan}, {van Straten}  \&
  {Manchester}}{{Hotan} et~al.}{2004}]{hotan_04}
{Hotan} A.~W.,  {van Straten} W.,   {Manchester} R.~N.,  2004, \mn@doi [\pasa]
  {10.1071/AS04022}, \href
  {https://ui.adsabs.harvard.edu/abs/2004PASA...21..302H} {21, 302}

\bibitem[\protect\citeauthoryear{Hulse \& Taylor}{Hulse \&
  Taylor}{1975}]{hulseandtaylor75}
Hulse R.~A.,  Taylor J.~H.,  1975, \mn@doi [Astrophys. J.] {10.1086/181708},
  195, L51

\bibitem[\protect\citeauthoryear{{Jankowski}, {van Straten}, {Keane}, {Bailes},
  {Barr}, {Johnston}  \& {Kerr}}{{Jankowski} et~al.}{2018}]{jankowski018}
{Jankowski} F.,  {van Straten} W.,  {Keane} E.~F.,  {Bailes} M.,  {Barr} E.~D.,
   {Johnston} S.,   {Kerr} M.,  2018, \mn@doi [\mnras] {10.1093/mnras/stx2476},
  \href {https://ui.adsabs.harvard.edu/abs/2018MNRAS.473.4436J} {473, 4436}

\bibitem[\protect\citeauthoryear{{Janssen}, {Stappers}, {Kramer}, {Nice},
  {Jessner}, {Cognard}  \& {Purver}}{{Janssen} et~al.}{2008}]{janssen08}
{Janssen} G.~H.,  {Stappers} B.~W.,  {Kramer} M.,  {Nice} D.~J.,  {Jessner} A.,
   {Cognard} I.,   {Purver} M.~B.,  2008, \mn@doi [\aap]
  {10.1051/0004-6361:200810076}, \href
  {https://ui.adsabs.harvard.edu/abs/2008A&A...490..753J} {490, 753}

\bibitem[\protect\citeauthoryear{{Johnston}, {Lyne}, {Manchester}, {Kniffen},
  {D'Amico}, {Lim}  \& {Ashworth}}{{Johnston} et~al.}{1992a}]{johnston92a}
{Johnston} S.,  {Lyne} A.~G.,  {Manchester} R.~N.,  {Kniffen} D.~A.,  {D'Amico}
  N.,  {Lim} J.,   {Ashworth} M.,  1992a, \mn@doi [\mnras]
  {10.1093/mnras/255.3.401}, \href
  {https://ui.adsabs.harvard.edu/abs/1992MNRAS.255..401J} {255, 401}

\bibitem[\protect\citeauthoryear{{Johnston}, {Manchester}, {Lyne}, {Bailes},
  {Kaspi}, {Qiao}  \& {D'Amico}}{{Johnston} et~al.}{1992b}]{johnstn_92}
{Johnston} S.,  {Manchester} R.~N.,  {Lyne} A.~G.,  {Bailes} M.,  {Kaspi}
  V.~M.,  {Qiao} G.,   {D'Amico} N.,  1992b, \mn@doi [\apjl] {10.1086/186300},
  \href {https://ui.adsabs.harvard.edu/abs/1992ApJ...387L..37J} {387, L37}

\bibitem[\protect\citeauthoryear{{Kaspi}, {Johnston}, {Bell}, {Manchester},
  {Bailes}, {Bessell}, {Lyne}  \& {D'Amico}}{{Kaspi} et~al.}{1994}]{kaspi_94}
{Kaspi} V.~M.,  {Johnston} S.,  {Bell} J.~F.,  {Manchester} R.~N.,  {Bailes}
  M.,  {Bessell} M.,  {Lyne} A.~G.,   {D'Amico} N.,  1994, \mn@doi [\apjl]
  {10.1086/187231}, \href
  {https://ui.adsabs.harvard.edu/abs/1994ApJ...423L..43K} {423, L43}

\bibitem[\protect\citeauthoryear{{Keith} et~al.,}{{Keith}
  et~al.}{2010}]{keith10}
{Keith} M.~J.,  et~al., 2010, \mn@doi [\mnras]
  {10.1111/j.1365-2966.2010.17325.x}, \href
  {https://ui.adsabs.harvard.edu/abs/2010MNRAS.409..619K} {409, 619}

\bibitem[\protect\citeauthoryear{{Keller} et~al.,}{{Keller}
  et~al.}{2007}]{skymapper_07}
{Keller} S.~C.,  et~al., 2007, \mn@doi [\pasa] {10.1071/AS07001}, \href
  {https://ui.adsabs.harvard.edu/abs/2007PASA...24....1K} {24, 1}

\bibitem[\protect\citeauthoryear{{Kramer} et~al.,}{{Kramer}
  et~al.}{2021a}]{kramer_21}
{Kramer} M.,  et~al., 2021a, \mn@doi [Physical Review X]
  {10.1103/PhysRevX.11.041050}, \href
  {https://ui.adsabs.harvard.edu/abs/2021PhRvX..11d1050K} {11, 041050}

\bibitem[\protect\citeauthoryear{{Kramer} et~al.,}{{Kramer}
  et~al.}{2021b}]{kramer_21b}
{Kramer} M.,  et~al., 2021b, \mn@doi [\mnras] {10.1093/mnras/stab375}, \href
  {https://ui.adsabs.harvard.edu/abs/2021MNRAS.504.2094K} {504, 2094}

\bibitem[\protect\citeauthoryear{Lattimer}{Lattimer}{2021}]{Lattimer:2021review}
Lattimer J.,  2021, \mn@doi [Annual Review of Nuclear and Particle Science]
  {10.1146/annurev-nucl-102419-124827}, 71, 433

\bibitem[\protect\citeauthoryear{{Lattimer} \& {Yahil}}{{Lattimer} \&
  {Yahil}}{1989}]{lattimer_and_yahil89}
{Lattimer} J.~M.,  {Yahil} A.,  1989, \mn@doi [\apj] {10.1086/167404}, \href
  {https://ui.adsabs.harvard.edu/abs/1989ApJ...340..426L} {340, 426}

\bibitem[\protect\citeauthoryear{{Lazarus} et~al.,}{{Lazarus}
  et~al.}{2015}]{lazarus15}
{Lazarus} P.,  et~al., 2015, \mn@doi [\apj] {10.1088/0004-637X/812/1/81}, \href
  {https://ui.adsabs.harvard.edu/abs/2015ApJ...812...81L} {812, 81}

\bibitem[\protect\citeauthoryear{{Lorimer}, {Bailes}, {McLaughlin}, {Narkevic}
  \& {Crawford}}{{Lorimer} et~al.}{2007}]{lorimer2007}
{Lorimer} D.~R.,  {Bailes} M.,  {McLaughlin} M.~A.,  {Narkevic} D.~J.,
  {Crawford} F.,  2007, \mn@doi [Science] {10.1126/science.1147532}, \href
  {https://ui.adsabs.harvard.edu/abs/2007Sci...318..777L} {318, 777}

\bibitem[\protect\citeauthoryear{{Lorimer} et~al.,}{{Lorimer}
  et~al.}{2015}]{lorimer_015}
{Lorimer} D.~R.,  et~al., 2015, \mn@doi [\mnras] {10.1093/mnras/stv804}, \href
  {https://ui.adsabs.harvard.edu/abs/2015MNRAS.450.2185L} {450, 2185}

\bibitem[\protect\citeauthoryear{{Lynch} et~al.,}{{Lynch}
  et~al.}{2018}]{lynch018}
{Lynch} R.~S.,  et~al., 2018, \mn@doi [\apj] {10.3847/1538-4357/aabf8a}, \href
  {https://ui.adsabs.harvard.edu/abs/2018ApJ...859...93L} {859, 93}

\bibitem[\protect\citeauthoryear{{Lyne} et~al.,}{{Lyne} et~al.}{2004}]{lyne_04}
{Lyne} A.~G.,  et~al., 2004, \mn@doi [Science] {10.1126/science.1094645}, \href
  {https://ui.adsabs.harvard.edu/abs/2004Sci...303.1153L} {303, 1153}

\bibitem[\protect\citeauthoryear{{Manchester} et~al.,}{{Manchester}
  et~al.}{2001}]{pmps01}
{Manchester} R.~N.,  et~al., 2001, \mn@doi [\mnras]
  {10.1046/j.1365-8711.2001.04751.x}, \href
  {https://ui.adsabs.harvard.edu/abs/2001MNRAS.328...17M} {328, 17}

\bibitem[\protect\citeauthoryear{{Manchester}, {Hobbs}, {Teoh}  \&
  {Hobbs}}{{Manchester} et~al.}{2005}]{psrcat05a}
{Manchester} R.~N.,  {Hobbs} G.~B.,  {Teoh} A.,   {Hobbs} M.,  2005, \mn@doi
  [\aj] {10.1086/428488}, \href
  {https://ui.adsabs.harvard.edu/abs/2005AJ....129.1993M} {129, 1993}

\bibitem[\protect\citeauthoryear{{Martinez} et~al.,}{{Martinez}
  et~al.}{2015}]{martinez015}
{Martinez} J.~G.,  et~al., 2015, \mn@doi [\apj] {10.1088/0004-637X/812/2/143},
  \href {https://ui.adsabs.harvard.edu/abs/2015ApJ...812..143M} {812, 143}

\bibitem[\protect\citeauthoryear{{Martinez} et~al.,}{{Martinez}
  et~al.}{2017}]{martinez017}
{Martinez} J.~G.,  et~al., 2017, \mn@doi [\apjl] {10.3847/2041-8213/aa9d87},
  \href {https://ui.adsabs.harvard.edu/abs/2017ApJ...851L..29M} {851, L29}

\bibitem[\protect\citeauthoryear{{Middleditch} \& {Kristian}}{{Middleditch} \&
  {Kristian}}{1984}]{middleditch84}
{Middleditch} J.,  {Kristian} J.,  1984, \mn@doi [\apj] {10.1086/161876}, \href
  {https://ui.adsabs.harvard.edu/abs/1984ApJ...279..157M} {279, 157}

\bibitem[\protect\citeauthoryear{{Morello} et~al.,}{{Morello}
  et~al.}{2019}]{morello18}
{Morello} V.,  et~al., 2019, \mn@doi [\mnras] {10.1093/mnras/sty3328}, \href
  {https://ui.adsabs.harvard.edu/abs/2019MNRAS.483.3673M} {483, 3673}

\bibitem[\protect\citeauthoryear{{Morello}, {Barr}, {Stappers}, {Keane}  \&
  {Lyne}}{{Morello} et~al.}{2020}]{morello_020}
{Morello} V.,  {Barr} E.~D.,  {Stappers} B.~W.,  {Keane} E.~F.,   {Lyne} A.~G.,
   2020, \mn@doi [\mnras] {10.1093/mnras/staa2291}, \href
  {https://ui.adsabs.harvard.edu/abs/2020MNRAS.497.4654M} {497, 4654}

\bibitem[\protect\citeauthoryear{{M{\"u}ller}, {Heger}, {Liptai}  \&
  {Cameron}}{{M{\"u}ller} et~al.}{2016}]{muller16}
{M{\"u}ller} B.,  {Heger} A.,  {Liptai} D.,   {Cameron} J.~B.,  2016, \mn@doi
  [\mnras] {10.1093/mnras/stw1083}, \href
  {https://ui.adsabs.harvard.edu/abs/2016MNRAS.460..742M} {460, 742}

\bibitem[\protect\citeauthoryear{{Ng} et~al.,}{{Ng} et~al.}{2015}]{cherry15}
{Ng} C.,  et~al., 2015, \mn@doi [\mnras] {10.1093/mnras/stv753}, \href
  {https://ui.adsabs.harvard.edu/abs/2015MNRAS.450.2922N} {450, 2922}

\bibitem[\protect\citeauthoryear{{Nomoto}}{{Nomoto}}{1984}]{nomoto84}
{Nomoto} K.,  1984, \mn@doi [\apj] {10.1086/161749}, \href
  {https://ui.adsabs.harvard.edu/abs/1984ApJ...277..791N} {277, 791}

\bibitem[\protect\citeauthoryear{{Onken} et~al.,}{{Onken}
  et~al.}{2019}]{skymapper_019}
{Onken} C.~A.,  et~al., 2019, \mn@doi [\pasa] {10.1017/pasa.2019.27}, \href
  {https://ui.adsabs.harvard.edu/abs/2019PASA...36...33O} {36, e033}

\bibitem[\protect\citeauthoryear{{Paczy{\'n}ski}}{{Paczy{\'n}ski}}{1971}]{Paczynski:1976IAUS}
{Paczy{\'n}ski} B.,  1971, \mn@doi [\araa]
  {10.1146/annurev.aa.09.090171.001151}, \href
  {https://ui.adsabs.harvard.edu/abs/1971ARA&A...9..183P} {9, 183}

\bibitem[\protect\citeauthoryear{{Parent} et~al.,}{{Parent}
  et~al.}{2018}]{parent_018}
{Parent} E.,  et~al., 2018, \mn@doi [\apj] {10.3847/1538-4357/aac5f0}, \href
  {https://ui.adsabs.harvard.edu/abs/2018ApJ...861...44P} {861, 44}

\bibitem[\protect\citeauthoryear{Peters}{Peters}{1964}]{peters64}
Peters P.~C.,  1964, \mn@doi [Phys. Rev.] {10.1103/PhysRev.136.B1224}, 136,
  B1224

\bibitem[\protect\citeauthoryear{{Phinney} \& {Kulkarni}}{{Phinney} \&
  {Kulkarni}}{1994}]{Phinney:1994ARA&A}
{Phinney} E.~S.,  {Kulkarni} S.~R.,  1994, \mn@doi [\araa]
  {10.1146/annurev.aa.32.090194.003111}, \href
  {https://ui.adsabs.harvard.edu/abs/1994ARA&A..32..591P} {32, 591}

\bibitem[\protect\citeauthoryear{{Podsiadlowski}, {Joss}  \&
  {Hsu}}{{Podsiadlowski} et~al.}{1992}]{Podsiadlowski_92}
{Podsiadlowski} P.,  {Joss} P.~C.,   {Hsu} J.~J.~L.,  1992, \mn@doi [\apj]
  {10.1086/171341}, \href
  {https://ui.adsabs.harvard.edu/abs/1992ApJ...391..246P} {391, 246}

\bibitem[\protect\citeauthoryear{{Podsiadlowski}, {Langer}, {Poelarends},
  {Rappaport}, {Heger}  \& {Pfahl}}{{Podsiadlowski}
  et~al.}{2004}]{podsiadlowski04}
{Podsiadlowski} P.,  {Langer} N.,  {Poelarends} A.~J.~T.,  {Rappaport} S.,
  {Heger} A.,   {Pfahl} E.,  2004, \mn@doi [\apj] {10.1086/421713}, \href
  {https://ui.adsabs.harvard.edu/abs/2004ApJ...612.1044P} {612, 1044}

\bibitem[\protect\citeauthoryear{{Shklovskii}}{{Shklovskii}}{1970}]{Shklovskii:1970SvA}
{Shklovskii} I.~S.,  1970, \sovast, \href
  {https://ui.adsabs.harvard.edu/abs/1970SvA....13..562S} {13, 562}

\bibitem[\protect\citeauthoryear{{Stappers} \& {Kramer}}{{Stappers} \&
  {Kramer}}{2016}]{trapum_16}
{Stappers} B.,  {Kramer} M.,  2016, in MeerKAT Science: On the Pathway to the
  SKA. p.~9

\bibitem[\protect\citeauthoryear{{Staveley-Smith} et~al.,}{{Staveley-Smith}
  et~al.}{1996}]{multibeam96}
{Staveley-Smith} L.,  et~al., 1996, \pasa, \href
  {https://ui.adsabs.harvard.edu/abs/1996PASA...13..243S} {13, 243}

\bibitem[\protect\citeauthoryear{{Stovall} et~al.,}{{Stovall}
  et~al.}{2014}]{gbncc_14}
{Stovall} K.,  et~al., 2014, \mn@doi [\apj] {10.1088/0004-637X/791/1/67}, \href
  {https://ui.adsabs.harvard.edu/abs/2014ApJ...791...67S} {791, 67}

\bibitem[\protect\citeauthoryear{{Stovall} et~al.,}{{Stovall}
  et~al.}{2018}]{stovall018}
{Stovall} K.,  et~al., 2018, \mn@doi [\apjl] {10.3847/2041-8213/aaad06}, \href
  {https://ui.adsabs.harvard.edu/abs/2018ApJ...854L..22S} {854, L22}

\bibitem[\protect\citeauthoryear{{Sutantyo}}{{Sutantyo}}{1978}]{sutantyo_78}
{Sutantyo} W.,  1978, \mn@doi [\apss] {10.1007/BF00639450}, \href
  {https://ui.adsabs.harvard.edu/abs/1978Ap&SS..54..479S} {54, 479}

\bibitem[\protect\citeauthoryear{{Suwa}, {Yoshida}, {Shibata}, {Umeda}  \&
  {Takahashi}}{{Suwa} et~al.}{2015}]{suwa015}
{Suwa} Y.,  {Yoshida} T.,  {Shibata} M.,  {Umeda} H.,   {Takahashi} K.,  2015,
  \mn@doi [\mnras] {10.1093/mnras/stv2195}, \href
  {https://ui.adsabs.harvard.edu/abs/2015MNRAS.454.3073S} {454, 3073}

\bibitem[\protect\citeauthoryear{{Swiggum} et~al.,}{{Swiggum}
  et~al.}{2015}]{swiggum015}
{Swiggum} J.~K.,  et~al., 2015, \mn@doi [\apj] {10.1088/0004-637X/805/2/156},
  \href {https://ui.adsabs.harvard.edu/abs/2015ApJ...805..156S} {805, 156}

\bibitem[\protect\citeauthoryear{{Tauris} \& {Takens}}{{Tauris} \&
  {Takens}}{1998}]{tauris_98}
{Tauris} T.~M.,  {Takens} R.~J.,  1998, \aap, \href
  {https://ui.adsabs.harvard.edu/abs/1998A&A...330.1047T} {330, 1047}

\bibitem[\protect\citeauthoryear{{Tauris}, {Langer}  \&
  {Podsiadlowski}}{{Tauris} et~al.}{2015}]{Tauris:2015xra}
{Tauris} T.~M.,  {Langer} N.,   {Podsiadlowski} P.,  2015, \mn@doi [\mnras]
  {10.1093/mnras/stv990}, \href
  {https://ui.adsabs.harvard.edu/abs/2015MNRAS.451.2123T} {451, 2123}

\bibitem[\protect\citeauthoryear{{Tauris} et~al.,}{{Tauris}
  et~al.}{2017}]{tauris17}
{Tauris} T.~M.,  et~al., 2017, \mn@doi [\apj] {10.3847/1538-4357/aa7e89}, \href
  {https://ui.adsabs.harvard.edu/abs/2017ApJ...846..170T} {846, 170}

\bibitem[\protect\citeauthoryear{{Taylor} \& {Weisberg}}{{Taylor} \&
  {Weisberg}}{1982a}]{Taylor:1982ApJ}
{Taylor} J.~H.,  {Weisberg} J.~M.,  1982a, \mn@doi [\apj] {10.1086/159690},
  \href {https://ui.adsabs.harvard.edu/abs/1982ApJ...253..908T} {253, 908}

\bibitem[\protect\citeauthoryear{{Taylor} \& {Weisberg}}{{Taylor} \&
  {Weisberg}}{1982b}]{taylor_82}
{Taylor} J.~H.,  {Weisberg} J.~M.,  1982b, \mn@doi [\apj] {10.1086/159690},
  \href {https://ui.adsabs.harvard.edu/abs/1982ApJ...253..908T} {253, 908}

\bibitem[\protect\citeauthoryear{{Thornton} et~al.,}{{Thornton}
  et~al.}{2013}]{thorton_013}
{Thornton} D.,  et~al., 2013, \mn@doi [Science] {10.1126/science.1236789},
  \href {https://ui.adsabs.harvard.edu/abs/2013Sci...341...53T} {341, 53}

\bibitem[\protect\citeauthoryear{Vigna-G\'omez et~al.}{Vigna-G\'omez
  et~al.}{2018}]{Vigna-Gomez:2018dza}
Vigna-G\'omez A.,  et~al., 2018, \mn@doi [Mon. Not. Roy. Astron. Soc.]
  {10.1093/mnras/sty2463}, 481, 4009

\bibitem[\protect\citeauthoryear{{Weisberg}, {Nice}  \& {Taylor}}{{Weisberg}
  et~al.}{2010}]{weisberg010}
{Weisberg} J.~M.,  {Nice} D.~J.,   {Taylor} J.~H.,  2010, \mn@doi [\apj]
  {10.1088/0004-637X/722/2/1030}, \href
  {https://ui.adsabs.harvard.edu/abs/2010ApJ...722.1030W} {722, 1030}

\bibitem[\protect\citeauthoryear{{Yao}, {Manchester}  \& {Wang}}{{Yao}
  et~al.}{2017}]{ymw16}
{Yao} J.~M.,  {Manchester} R.~N.,   {Wang} N.,  2017, \mn@doi [\apj]
  {10.3847/1538-4357/835/1/29}, \href
  {https://ui.adsabs.harvard.edu/abs/2017ApJ...835...29Y} {835, 29}

\bibitem[\protect\citeauthoryear{{van Leeuwen} et~al.,}{{van Leeuwen}
  et~al.}{2015}]{leeuwen015}
{van Leeuwen} J.,  et~al., 2015, \mn@doi [\apj] {10.1088/0004-637X/798/2/118},
  \href {https://ui.adsabs.harvard.edu/abs/2015ApJ...798..118V} {798, 118}

\bibitem[\protect\citeauthoryear{{van den Heuvel}}{{van den
  Heuvel}}{2019}]{vanden_019}
{van den Heuvel} E. P.~J.,  2019, \mn@doi [IAU Symposium]
  {10.1017/S1743921319001315}, \href
  {https://ui.adsabs.harvard.edu/abs/2019IAUS..346....1V} {346, 1}

\bibitem[\protect\citeauthoryear{Özel \& Freire}{Özel \&
  Freire}{2016}]{ozel_016}
Özel F.,  Freire P.,  2016, \mn@doi [Annual Review of Astronomy and
  Astrophysics] {10.1146/annurev-astro-081915-023322}, 54, 401–440

\makeatother
\end{thebibliography}




\bsp	
\label{lastpage}

\end{document}